\documentclass{aa}  
\usepackage{graphicx}
\usepackage{txfonts}
\usepackage{hyperref}
\usepackage{float}
\usepackage[separate-uncertainty=true]{siunitx}
\usepackage{booktabs}
\usepackage{lipsum} 

\newcommand{\expected}[1]{\langle #1 \rangle}
\begin{document} 
   \title{Unusual periodic modulation in the radio emission of the methane dwarf binary WISEP J101905.63+652954.2}
   \titlerunning{Unusual periodicity in J1019+65}

   \author{T. W. H. Yiu
          \inst{1,2}\fnmsep\thanks{Corresponding email: yiu@astron.nl}
          \and
          H. K. Vedantham\inst{1,2}
          \and
          J. R. Callingham\inst{1,3}
          \and
          T. W. Shimwell\inst{1,4}
          }

   \institute{ASTRON, The Netherlands Institute for Radio Astronomy, Oude Hoogeveensedijk 4, Dwingeloo, 7991 PD, The Netherlands
         \and
             Kapteyn Astronomical Institute, University of Groningen, PO Box 72, 97200 AB, Groningen, The Netherlands
         \and
            Anton Pannenkoek Institute for Astronomy, University of Amsterdam, Science Park 904, 1098\,XH, Amsterdam, The Netherlands
         \and
             Leiden Observatory, Leiden University, PO Box 9513, 2300 RA Leiden, The Netherlands
             }

   \date{Received 28 March 2025; accepted 21 October 2025}

\abstract{Brown dwarfs display Jupiter-like auroral phenomena, such as rotationally modulated electron cyclotron maser radio emission. Radio observations of cyclotron maser emission can be used to measure their magnetic field strength, topology, and to deduce the presence of magnetically interacting exoplanets.
Observations of the coldest brown dwarfs (spectral types T and Y) are especially intriguing, as their magnetospheric phenomena could closely resemble those of gas-giant exoplanets.
Here we report observations made over ten epochs, amounting to 44 hours, of WISEP J101905.63+652954.2 (J1019+65 hereinafter) using the LOFAR telescope between 120 and 168 MHz.
J1019+65 is a methane dwarf binary (T5.5\,$+$\,T7) whose radio emission was originally detected in a single-epoch LOFAR observation to be highly circular polarised and rotationally modulated at $\approx 3$\,h.
Unexpectedly, our long-term monitoring reveals an additional periodic signature at $\approx 0.787$\,h. We consider several explanations for the second period and suggest that it could be the rotationally modulated emission of the second brown dwarf in the binary, although follow-up infrared observations are necessary to confirm this hypothesis. 
In addition, the data also allow us to statistically estimate the duty cycle and observed radio-loud fraction of the 120--168\,MHz cyclotron emission from methane dwarfs to be $\expected{D} = 0.030^{+0.034}_{-0.030}$ and $F^{'}_{\rm radio} = 0.088^{+0.168}_{-0.088}$ respectively.
}

   \maketitle

\section{Introduction}
\label{sec:intro}

Brown dwarfs are known to exhibit rotationally modulated coherent radio bursts \citep{hallinan2006, hallinan2007, hallinan2008, berger2009, route2012, williams2015, route2016, kao2016, kao2018, harish_j1019, rose2023}.
These radio bursts are generally attributed to electron cyclotron maser instability (ECMI; \citealp{wu1979, treumann2006, zarka_bible2007, hallinan2008}) from non-thermal charges trapped in the magnetosphere, which is the principal mechanism that drives the auroral\footnote{The term ``auroral'' refers to the fact that the same population of electrons strike the planet's atmosphere, leading to auroral phenomena on the planet.} radio emission in Jupiter and other magnetised planets in the Solar System \citep{zarka1998}.

Owing to remote and in-situ observations of Jovian auroral decametric emission (DAM; see \citealp{marques2017} and references therein), the basic properties of electron cyclotron maser (ECM) radio emission stemming from a magnetopshere are well understood:
ECM emission occurs at the local cyclotron frequency and is beamed along the surface of a cone. The opening angle of the cone is wide and its axis parallel to the magnetic field line. Therefore, the aggregate radio emission from different magnetospheric regions typically appears as a broadband signal that is modulated by the rotation rate of the magnetosphere (akin to a lighthouse; \citealp{joe2024}). In the case of Jupiter, its ECM emission is modulated at around 10 hours, which is the rotation period of the Jovian magnetosphere. This measurement technique has provided the most accurate measurement of the rotation period of gas giants in the Solar System (e.g. \citealp{higgins1997, carbary2013}). It has also been successful in determining the rotation period of brown dwarfs down to spectral class T (i.e. methane dwarfs) \citep{joe2024}.

Jovian ECM emission is also modulated in a second period corresponding to the orbital period of Io \citep{marques2017}. This component of the emission is induced by the interaction between the conductive ionosphere of Io and the Jovian magnetic field \citep{zarka1998}. Because the electrons involved in this component only exist along the flux tube connecting Jupiter and Io, beaming ensures that the emission reaches the observer only when Io is at certain longitudes in its orbit. Identifying a second periodicity in brown dwarfs could analogously lead to the identification of exoplanets. Exoplanets around cold methane dwarfs likely exist but have eluded discovery so far \citep{kipping2009,kipping2012}.

The rotation periods of radio-loud brown dwarfs are usually in the range of a few to several hours measured via their radio light curves \citep{kao2016,kao2018,williams2015,route2012,route2016}. The rotational periodic modulation can therefore be detected with a modest amount of telescope exposure. However, the orbital periods of putative exoplanets around brown dwarfs could be in the range of days. Detecting the second periodicity therefore requires significant telescope time. 

\cite{harish_j1019} recently discovered a new radio-loud brown dwarf system. Located at $\num{23.3 \pm 1.0}$\,pc \citep{kirkpatrick2019}, WISEP J101905.63+652954.2 ($=$~2MASS J10190575+6529526~$=$~WISEA J101905.61+652954.0; hereinafter J1019+65) consists of two brown dwarfs in a binary with spectral types T$5.5 \pm 0.5$ and T$7.0 \pm 0.5$ \citep{harish_j1019}. The two brown dwarfs have a projected separation of $423.0 \pm 1.6$\,mas which corresponds to $\num{9.9 \pm 0.4}$\,AU \citep{harish_j1019}. J1019+65 showed circularly polarised radio pulsations at 144\,MHz using data from the LOFAR Two-metre Sky Survey (LoTSS; \citealp{lotss}).
This marks the first methane dwarf binary to be detected in the radio wavelength. Using the Lomb-Scargle (LS) periodogram \citep{lomb1976, scargle1982}, \cite{harish_j1019} also revealed a periodic signature of $\approx \qty{3}{hours}$ from the LOFAR-band light curve of J1019+65. \cite{harish_j1019} interpreted this radio-derived three-hour period as the rotation period of the radio emitter in J1019+65. However, since LoTSS has an absolute astrometric precision of 0.5\arcsec (defined by the Gaussian-equivalent standard deviation; \citealp{lotss_dr2}), it is not possible to discern which of the two brown dwarfs in J1019+65 is the radio emitter (or perhaps both are) with the LoTSS data alone due to insufficient angular resolution.

Despite the complication of identifying the radio-loud component in J1019+65, it is a well-suited system for monitoring for the following reasons: Since LoTSS is an untargeted all-sky survey, the radio detection of J1019+65 does not suffer from selection bias which is inherent to targeted observations. Thus, by studying J1019+65 further, we can learn more about the population of radio-loud brown dwarfs using J1019+65 as an archetype. Furthermore, unlike isolated brown dwarfs, the binary nature of J1019+65 will allow for an independent mass constraint using future infrared observations \citep{harish_j1019}. 

In this work, we present our observational campaign of J1019+65 with the Low-Frequency Array (LOFAR). The campaign comprises 44 hours of new radio observations plus the original 8-hour LoTSS data, spanning 6 years (2017--2023), making it one of the most intensive long-term monitoring of a brown dwarf system in radio wavelength to date.
The paper is structured as follows. In Sect.~\ref{sec:obs}, we present the details regarding the follow-up observations of J1019+65 carried out by LOFAR, which also include the description of the data reduction procedure.  We discuss our results in Sect.~\ref{sec:results} and conclude in Sect.~\ref{sec:conclusion}.


\setlength{\tabcolsep}{15pt}
\renewcommand{\arraystretch}{1.5}

\begin{table}
    \caption{Long-term radio monitoring of J1019+65 using LOFAR between 120 and 168 MHz.}
    \centering
    \begin{tabular}{c c c}
        \toprule
        Start of Epoch & Exposure time & $\overline{\sigma}_{\rm V, 4min}$ \\
        (UTC) &  (hours) & (\unit{\mu Jy}) \\
        \midrule
        2021-02-23T19:01:20.0  & 8   &  476         \\
        2022-12-01T03:06:00.0  & 4   &  542         \\
        2023-01-15T00:11:00.0  & 4   &  658         \\
        2023-01-24T21:41:00.0  & 4   &  556        \\
        2023-02-12T21:41:00.0  & 4   &  568         \\
        2023-02-14T22:11:00.0  & 4   &  572         \\
        2023-03-05T20:11:00.0  & 4   &  746         \\
        2023-03-08T18:11:00.0  & 4   &  698         \\
        2023-03-11T23:11:00.0  & 4   &  833         \\
        2023-03-20T20:11:00.0  & 4   &  616         \\
        \bottomrule
    \end{tabular}
    \tablefoot{$\sigma_{\rm V, 4min}$ represents the local root-mean-square (RMS) noise around J1019+65 averaged across all Stokes V 4min-cadence images (each synthesised over the full bandwidth). The date and time formats follow ISO 8601. The exposure time includes overheads. 
    }
    \label{tab:observations}
\end{table}

\section{Observations and data reduction}
\label{sec:obs}

Since the discovery of J1019+65's radio emission in the LOFAR Two Metre Sky Survey data release 2 (LoTSS DR2) exposure, we have acquired 10 new follow-up observations with LOFAR as part of projects \texttt{LC15\_019} \& \texttt{LT16\_013} (details in Table in Table~\ref{tab:observations}). Except for the first epoch (2021) of our follow-up observation (project code: \texttt{LC15\_019}) which has the same exposure time (8 hours) as the LoTSS discovery field, all other LOFAR observations (project code: \texttt{LT16\_013}) lasted 4 hours each to ensure that we covered the full three-hour rotation (with a 2$\sigma$ uncertainty range of $2.6-3.8$ hours; \citealp{harish_j1019}) of J1019+65 in each observation.

We processed the LOFAR dataset for each epoch using the standard LOFAR Initial Calibration (LINC) pipeline\footnote{\url{https://git.astron.nl/RD/LINC}} to flag interference and correct for direction-independent errors. 
See \cite{linc_degasperin2019} for more details of these effects and the calibration strategy used in the LINC pipeline.
Afterwards, we processed each dataset using the direction-dependent self-calibration and imaging pipeline (\texttt{DDF-pipeline}\footnote{\url{https://github.com/mhardcastle/ddf-pipeline}}).
The \texttt{DDF-pipeline} is described by \cite{lotss_dr2} and \cite{tasse2021}. The procedure also involved extraction and self-calibration to remove other sources in the field before self-calibrating on the target field, as described by \cite{vanweeren2021}.
Finally, we synthesised the full-exposure Stokes I and V images of J1019+65 for each LOFAR epoch using \texttt{wsclean} with Briggs weighting \citep{briggs1995}. We used robustness parameters of $-0.5$ and $0.0$ for Stokes I and Stokes V images, respectively.

The post-processing analysis for all datasets follows the procedure described by \cite{harish_elegast2020, harish_j1019}. 
For each calibrated dataset, we first used \texttt{wsclean} \citep{wsclean2014} to create Stokes V dirty snapshot images synthesised over the full bandwidth ($120-168$\,MHz) at a cadence of 4 minutes after phase-shifting the visibilities to the proper-motion-corrected sky location of J1019+65. We used the ephemeris of \cite{kirkpatrick2019} to account for proper motion.
Furthermore, we split the data into three spectral bins ($120-136$\,MHz, $136-152$\,MHz, and $152-168$\,MHz) to check for any spectral evolution in the light curves.
We then produced light curves for each epoch and spectral bin by extracting the value of the flux density at J1019+65 from the 4min images.
For each light curve, we computed the Lomb-Scargle (LS) periodogram using the \texttt{astropy} implementation \citep{astropy:2013, astropy:2018, astropy:2022} to determine if there were any strong peaks in the LS power spectrum of J1019+65. To quantify the significance of a periodogram peak, we computed its false-alarm probability (FAP) using the approach of \cite{baluev2008} that was included in the \texttt{astropy} implementation. 
The FAP represents the probability that a light curve with no periodic component would lead to a peak of a similar height as the one detected \citep{vanderplas2018}.
Figure~\ref{fig-app:lofar-2017-multi-lcls} in the Appendix shows the post-processed light curve and periodogram from the original LoTSS 2017 epoch and serves as a reference for our follow-up radio observations.

\begin{figure*}
    \centering
    \includegraphics[width=0.9\textwidth]{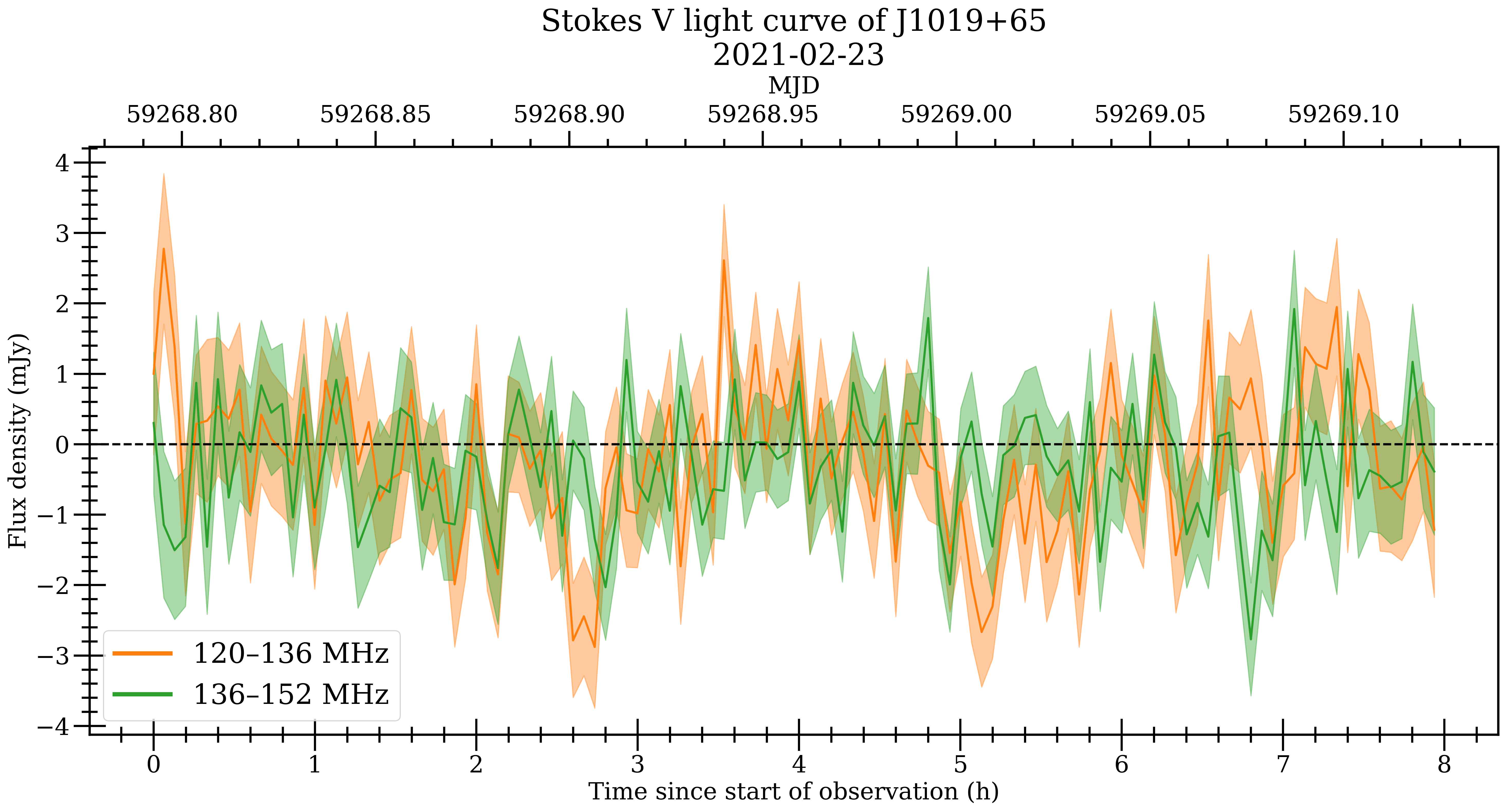}
    \includegraphics[width=0.9\textwidth]{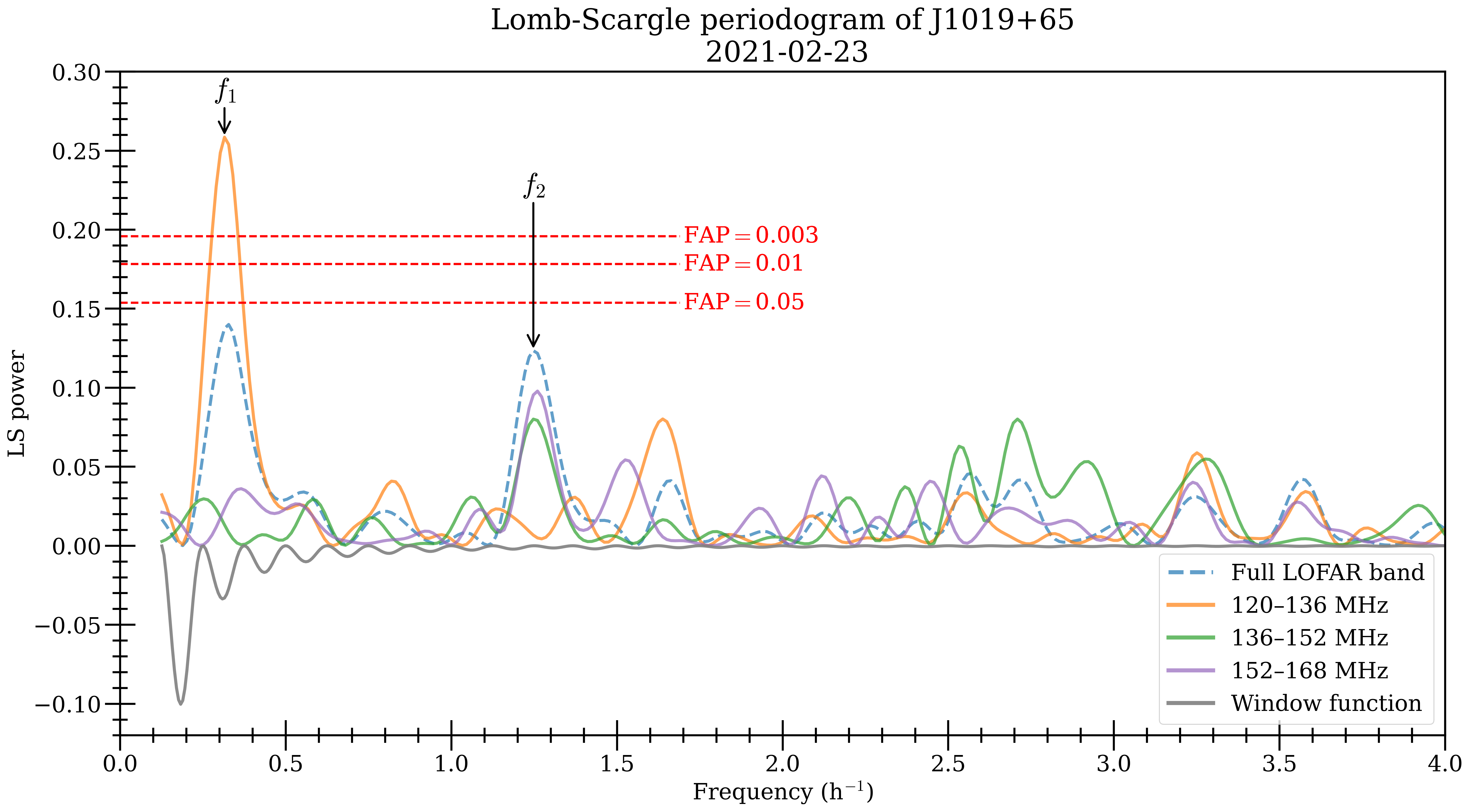}
    \caption{LOFAR observation of J1019+65 carried out on 2021-02-33. 
    Top panel: Stokes V light curves with a cadence of 4 minutes at different frequencies. The shaded region represents $\pm 1 \sigma$ uncertainty. The light curve at 152--168\,MHz is omitted for the sake of clarity. The top axis represents the time of observation in Modified Julian Dates (MJD). The black dashed line at 0\,mJy is drawn for clarity.
    Bottom panel: Lomb-Scargle periodogram of the radio light curves. The three red dashed lines represent the necessary LS power (i.e. peak height) to achieve a false-alarm probability (FAP) of \qty{5}{\percent}, \qty{1}{\percent}, and \qty{0.3}{\percent}. The grey curve represents the (negative) LS periodogram of the window function, which is a light curve with the same timestamps as the original curve, but the flux density of each data point is replaced with unity (i.e. a flat light curve).
    The two relevant frequencies $f_1 = 0.315 \pm 0.013\,\unit{h^{-1}}$ and $f_2 = 1.247 \pm 0.013\,\unit{h^{-1}}$ are indicated by the black arrows in the Lomb-Scargle periodogram. The 120--136\,MHz (orange) peak at $f_1$ is consistent with the original frequency discovered by \cite{harish_j1019}, while the relevancy of the $f_2$ is discussed in Sects.~\ref{sec:results-fidelity}. The frequency uncertainties were computed using Eq.~52 by \cite{vanderplas2018}.
    }
    \label{fig:lofar-2021-multi-lcls}
\end{figure*}

\subsection{Single-epoch analysis}
\label{sec:results-single}
We individually inspected all the Stokes V light curves of the radio observations. None of the 10 LOFAR follow-up epochs shows radio bursts in four-minute cadence data above a threshold of 5$\sigma$. The lack of new short bursts in the follow-up observations is surprising but not unexpected because ECMI, a coherent emission mechanism, is intrinsically highly variable. For example, the occurrence probability of Jovian DAM emission (as a function of Central Meridian Longitude in Jupiter’s System III coordinates, or CML for short) only reaches a maximum of around \qty{30}{\percent} at certain values of CML \citep{marques2017}.
As the occurrence probability of ECM emission from known brown dwarf emitters remains largely unconstrained, it is entirely possible that J1019+65 was simply not producing sufficiently bright and singular ECM burst for detection during most of our observations.

In terms of periodicity, only the 8-hour follow-up observation (2021-02-23) shows a significant detection (FAP $\ll 0.05$). The rest are only 4 hours long and thus cannot reliably measure the 3-hour rotation period of J1019+65 in a single exposure.
Figure~\ref{fig:lofar-2021-multi-lcls} shows the radio light curve of the 2021-02-23 epoch and the corresponding LS periodogram.
Although we did not detect a single radio pulsation that is as strong as the original pulse from LoTSS DR2, the periodogram clearly detects the original period of $\approx 3$\,h. We shall hereafter refer to this frequency as $f_1$ as represented in Fig.~\ref{fig:lofar-2021-multi-lcls}.
However, in contrast to the original 2017-03-08 epoch, the $f_1$ peak is not significant for the full-bandwidth light curve (blue); the $f_1$ peak is much more significant (FAP $\ll 0.003$) at 120--136\,MHz (orange), while the other two spectral bins (136--152\,MHz and 152--168\,MHz; green and purple) did not contribute to this peak. This suggests that there is some spectral shape evolution with time in the rotationally modulated emission.

The other 9 LOFAR single-epoch light curves and their corresponding LS periodograms can be found in Figs.~\ref{fig-app:lofar-3-obs} and \ref{fig-app:lofar-6-obs} in the Appendix.

\begin{figure*}
    \centering
    \includegraphics[width=0.95\textwidth]{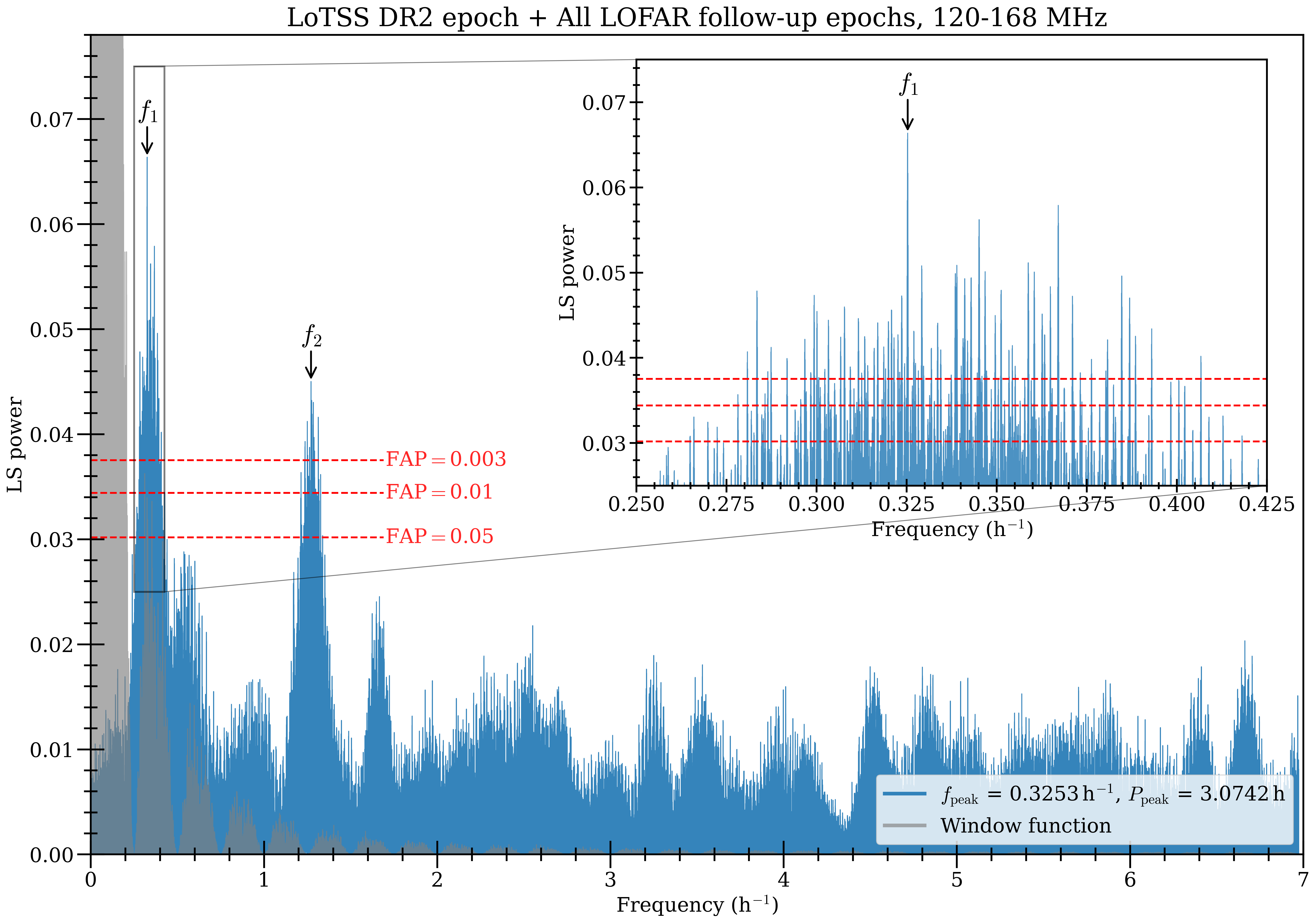}
    \caption{Lomb-Scargle periodograms of the LOFAR cross-epoch radio light curve, which includes data from both the original LoTSS DR2 epoch and our 10 LOFAR follow-up epochs.
    The spectrum reveals the peak at $f_1 \approx 0.325\,\unit{h^{-1}}$ originally discovered by \cite{harish_j1019}.
    Another significant peak not previously discovered is detected at $f_2 \approx 1.271\,\unit{h^{-1}}$, with a false alarm probability (FAP) of less than \qty{0.3}{\percent}.
    The locations of these two peaks are indicated by the black arrows. 
    The three red dashed lines represent the FAP of different values. The grey curve represents the LS periodogram of the window function, which is a light curve with the same timestamps as the original curve, but the flux density of each data point is replaced by unity (i.e. a flat light curve). 
    The inset plot in each panel shows a zoomed-in version of the observed J1019+65 power spectrum in the vicinity of the tallest LS peak, with the same three FAP red dashed lines.
    }
    \label{fig:cross-epoch-ls}
\end{figure*}

\subsection{LOFAR cross-epoch analysis}
\label{sec:results-cross}

We also concatenated all 10 LOFAR follow-up light curves plus the original LoTSS DR2 light curve of J1019+65 to create a cross-epoch light curve that spans over 6 years (from 2017-03-08 to 2023-03-20), and computed its LS periodogram, as shown in Fig.~\ref{fig:cross-epoch-ls}. 
Note that one can see the peculiar fine-scale structure around the spectral peaks in the zoomed-in inset figure. The structure stems from the observational pattern (i.e. the window function) because the observed spectrum is a convolution of the true spectrum and the window spectrum (grey), which has a complex structure due to a combination of long temporal baseline ($>2$~year) of our data and the 4min ``sampling'' interval of the light curve. See \cite{vanderplas2018} for more details regarding the various effects of the window function on the observed power spectrum in the LS periodogram.

\subsubsection{New periodic signature}
\label{sec:results-new-period}

Fig.~\ref{fig:cross-epoch-ls} shows that the original periodogram peak originally discovered by \cite{harish_j1019} is present at $f_1 \approx 0.325\,\unit{h^{-1}}$ as expected. 
In addition, we discovered a new periodicity in our cross-epoch light curve that was not seen by \cite{harish_j1019}: There exists a statistically significant peak at $f_2 \approx 1.271\,\unit{hour^{-1}}$ which corresponds to a period $P_2 \approx 0.787$\,h. The FAP of this peak is less than \qty{0.3}{\percent}.

Note that we do not report the formal uncertainty of the measured frequency in this periodogram analysis due to the reasons discussed by \cite{vanderplas2018}. Briefly, the frequency uncertainty is defined by the full width at half maximum (FWHM) of the periodogram peak which is equal to the inverse of the temporal baseline. However, such an assignment of uncertainty is fraught for sparse sampling over a large temporal baseline -- 6 years long in our case. This is because the presence of numerous adjacent peaks in the window function makes it impossible to identify the true peak. Therefore, a more pragmatic choice for the uncertainties of $f_1$ and $f_2$ in this cross-epoch analysis would be the FWHM of the envelope of the first peak of the window function, which is approximately 0.2\,\unit{h^{-1}} shown in Fig.~\ref{fig:cross-epoch-ls}. Using this definition of FWHM and Eq.~52 by \cite{vanderplas2018}, the uncertainties in $f_1$ and $f_2$ were computed: $f_1 = 0.325 \pm 0.006\,\unit{h^{-1}}$ and $f_2 = 1.271 \pm 0.006\,\unit{h^{-1}}$.

\subsubsection{Fidelity of $f_1$ and $f_2$}
\label{sec:results-fidelity}

\begin{figure}
    \centering
    \resizebox{\hsize}{!}{\includegraphics[width=1.0\textwidth]{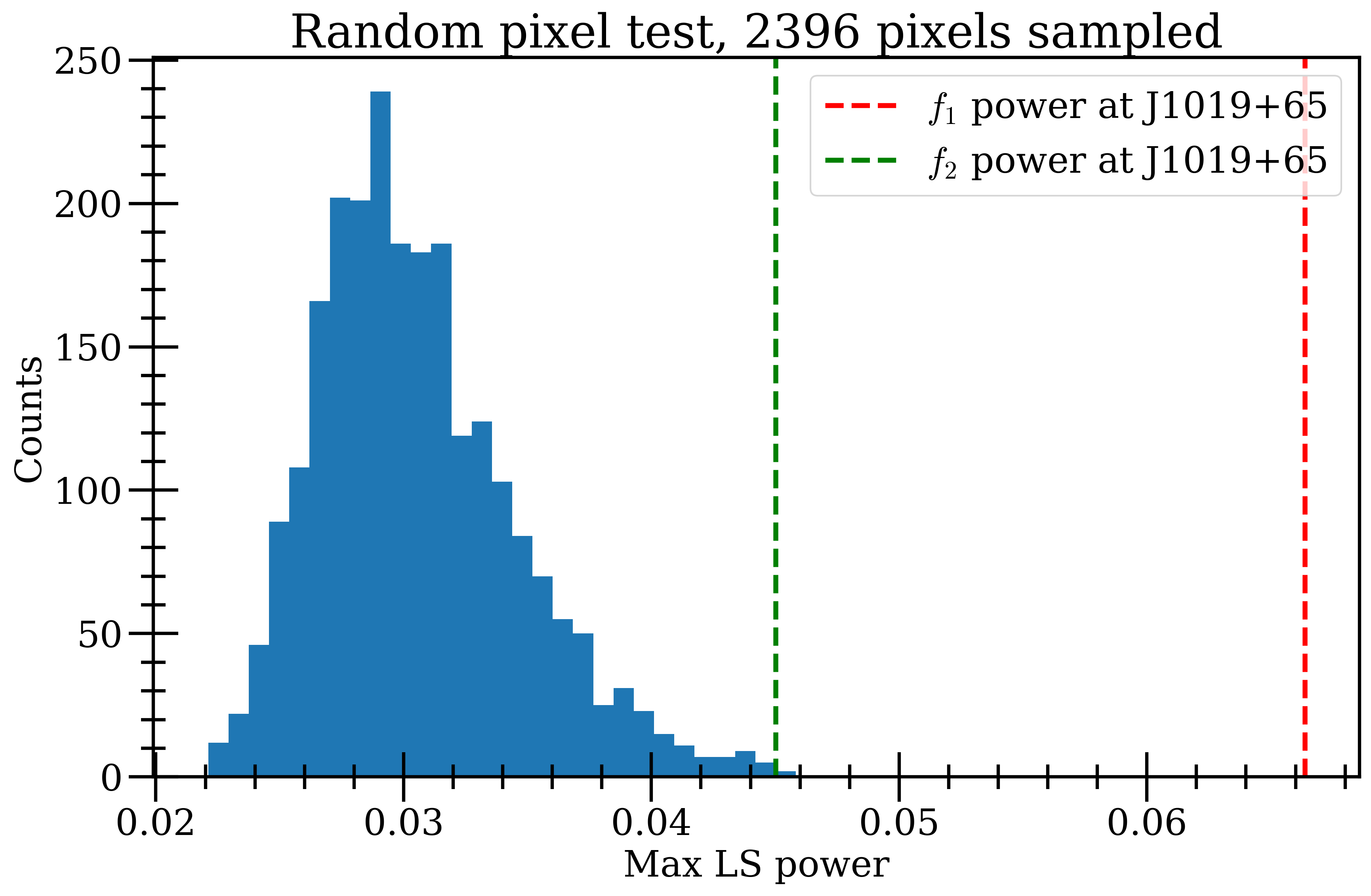}}
    \caption{Histogram of peak LS power calculated from light curves at random sky locations in the vicinity of J1019+65.}
    \label{fig:random-pixel-test}
\end{figure}

Despite a sufficiently small value of FAP for the significant frequencies, a periodic signature in an LS periodogram could still be spurious if they are caused by e.g. image artefacts such as sidelobes from nearby bright sources. To test the fidelity of the two detected periodic signatures of the radio light curve of J1019+65 at $P_1 \approx 3.077 \pm 0.057 $\,h and $P_2 \approx 0.787 \pm 0.006 $\,h, we used the following method:
We first pick a random\footnote{Due to the effect of point spread function, we avoid sampling pixels too close ($\lesssim 6\arcsec$) to the sky location of J1019+65 and of random pixels already sampled. However, we also exclusively sample pixels in a $5\arcmin$ radius region centred at J1019+65 to ensure that the local RMS noise around the sampled pixels is similar to that around J1019+65.} sky location in the field of J1019+65 and sample the corresponding pixel in the LOFAR Stokes V image, which involves computing the cross-epoch light curve and the LS periodogram of this random pixel. The LS power (i.e. height) of the most significant (i.e. highest) peak in this periodogram is then recorded, before we repeat this trial many more times in order to get an ensemble of the maximum peak height of each random-pixel power spectrum.
The idea behind this test is that if the periodicities detected in J1019+65's light curve are, in fact, caused by sidelobes of nearby bright sources, other sky locations in the field should also show spurious periodic signatures of similar significance in their periodograms. Therefore, we performed 2396 trials (each on a different pixel) and determined that $<\qty{0.3}{\percent}$ of all random-pixel power spectra has a peak that exceeds the height of the $f_2$ peak, and $\ll \qty{0.3}{\percent}$ of the $f_1$ peak, shown in Fig.~\ref{fig:random-pixel-test}. We can thus empirically eliminate the possibility that the $f_1$ and $f_2$ peaks were caused by image artefacts.

\begin{figure}
    \centering
    \resizebox{\hsize}{!}{\includegraphics[width=1.0\textwidth]{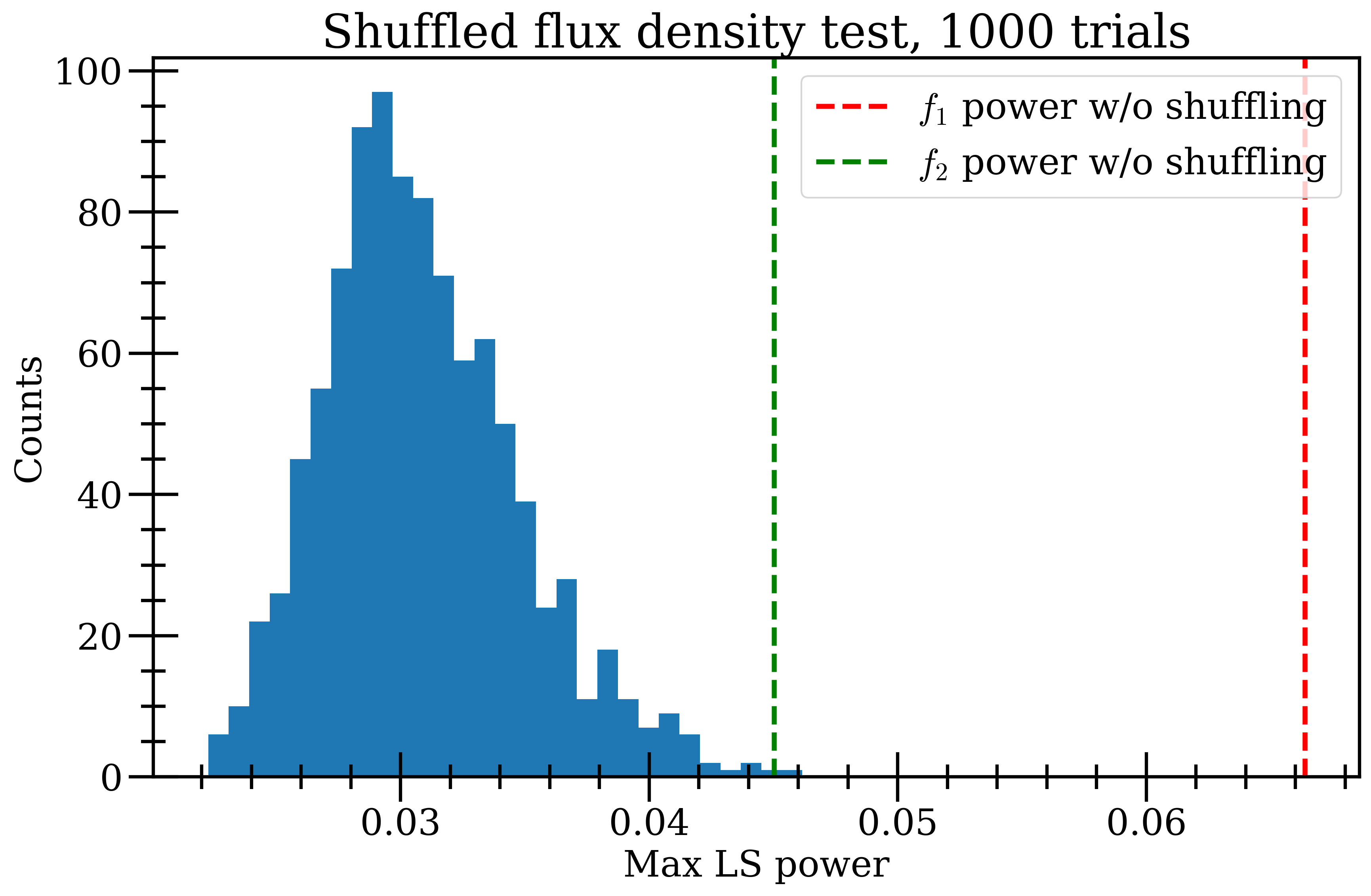}}
    \caption{Histogram of peak LS power calculated by randomly shuffling the measured light curve values while keeping the same time sampling window.}
    \label{fig:shuffled-flux-test}
\end{figure}

Another test of the fidelity of $f_1$ and $f_2$ is to randomly shuffle the data points of the original light curve and compute the LS periodogram of the shuffled light curve. The maximum peak height of the power spectrum was then recorded each time we shuffled the data points. We performed 1000 trials and determined that $<\qty{0.3}{\percent}$ of all flux-density-shuffled power spectra has a peak that exceeds the height of the $f_2$ peak, and $\ll \qty{0.3}{\percent}$ of the $f_1$ peak, shown in Fig.~\ref{fig:shuffled-flux-test}.

As shown in Fig.~\ref{fig-app:cross-epoch-ls-multi}, we can see that unlike the $f_1$ signature which is only present in the 120--136\,MHz, the $f_2$ signature shows up still at higher frequencies of 136--168 MHz. This implies a difference between the spectral shape of the radio emission that is responsible for the $f_1$ and $f_2$ signatures.

\begin{figure}
    \centering
    \resizebox{\hsize}{!}{\includegraphics[width=1.0\textwidth]{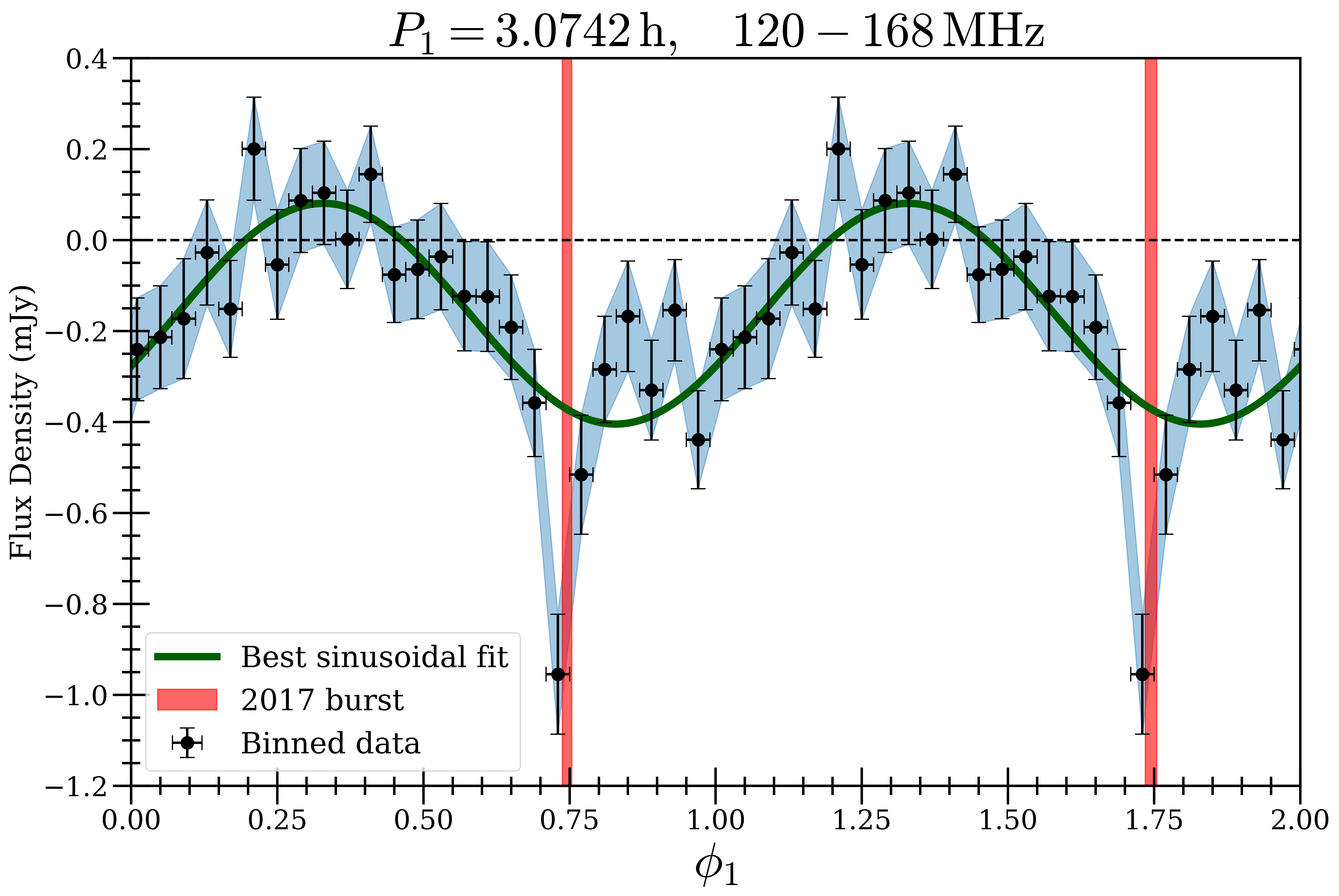}}
    \caption{Stokes V cross-epoch (2017--2023) light curve of J1019+65 phase-folded at a period of $P_1=3.0742$\,h corresponding to the $f_1$ peak seen in Fig.~\ref{fig:cross-epoch-ls}. The light curve is binned at intervals of 0.04 in phase and one full period corresponds to a phase interval of unity. The choice of starting phase is arbitrary. The green line represents the best sinusoidal fit using the computed Lomb-Scargle model. The J1019+65 radio burst shown in Fig.~\ref{fig-app:lofar-2017-multi-lcls} is also plotted as a red line (with its width representing the duration of the burst) to indicate at which phase relative to the phase-folded light curves does it correspond to. Note that the burst shown in red only occurs in one epoch out of the whole dataset. The black dashed line at 0\,mJy is drawn for clarity.}
    \label{fig:phasefolded_lc_3h}
\end{figure}

\begin{figure}
    \centering
    \resizebox{\hsize}{!}{\includegraphics[width=1.0\textwidth]{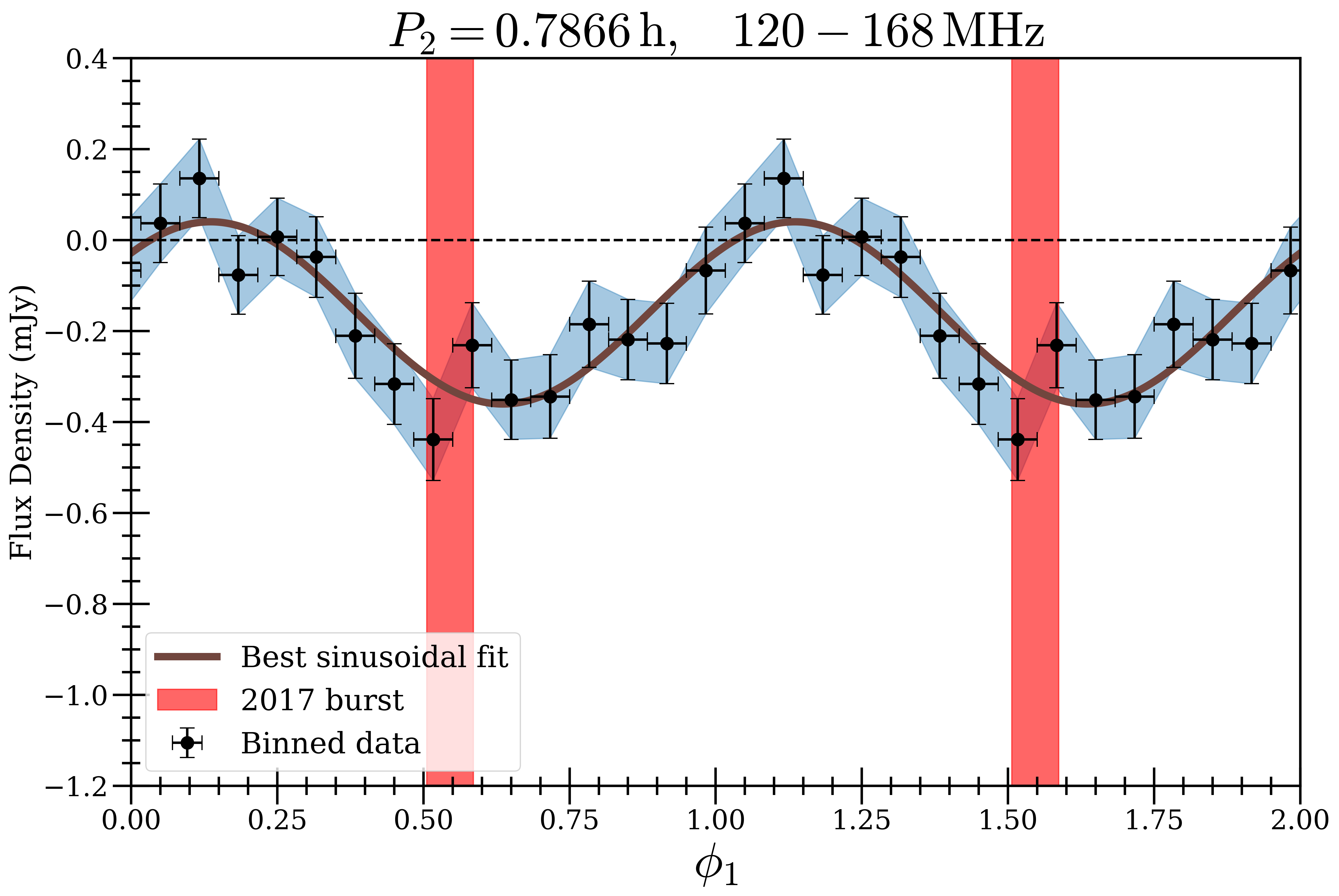}}
    \caption{Same as Fig~\ref{fig:phasefolded_lc_3h}, but instead with a shorter period of $P_2=0.7866$\,h corresponding to the $f_2$ peak seen in Fig.~\ref{fig:cross-epoch-ls}.}
    \label{fig:phasefolded_lc_1h}
\end{figure}

\subsubsection{Phase-folded light curve \& stacked image}
\label{sec:results-phasefold-stacked}

Figures~\ref{fig:phasefolded_lc_3h} and \ref{fig:phasefolded_lc_1h} show the cross-epoch light curve when phase-folded with each of two significant periods -- $P_1=3.07$\,h and $P_2=0.79$\,h respectively -- derived from the $f_1$ and $f_2$ peaks discussed in Sect.~\ref{sec:results-new-period}. 
From the figures, we can see that both phase-folded light curves are mostly below the zero line, implying that J1019+65 was producing a significant amount of phase-averaged (negative) Stokes V flux density in sub-mJy scales. Since a single normal LOFAR observation can only reach a $1\sigma$ sensitivity of $\approx 0.2$\,mJy in an 8-hour exposure \citep{v-lotss}, a single epoch of observation would not be able to reveal the periodic emission. However, the intrinsic variability of the periodic emission means that it can occasionally be detected in single-epoch observations. Thus, the detection of this persistent emission -- derived from LOFAR observations that comprise 52 hours worth of data -- is entirely consistent with the non-detections in the 4h or 8h observation of individual LOFAR epoch but with single-epoch detections in the 2017-03-08 and 2021-02-03 epochs.

More importantly, it appears that this emission was detected mainly within a certain phase $\phi$ range. For example, as shown in Fig.~\ref{fig:phasefolded_lc_3h}, J1019+65 has significantly more negative Stokes V emission, both on average (with a weighted mean $\overline{S}_{\rm V} = \num{-0.136 \pm 0.029}$\,mJy) and when $\phi_1$ takes a value between 0.3 and 0.8 approximately, reaching its peak at $\phi_1 \approx 0.55$. We see the same phenomena in Fig.~\ref{fig:phasefolded_lc_1h}. However, note that the quasi-sinusoidal profile of the data in Figs.~\ref{fig:phasefolded_lc_3h} and \ref{fig:phasefolded_lc_1h} should be interpreted with caution, as the LS algorithm specifically finds periods that yield a sinusoidal phase-folded light curve. In other words, it may be possible to find a period very close to the LS peak that yields a different (non-sinusoidal) period signature.

\begin{figure}
    \centering
    \resizebox{\hsize}{!}{\includegraphics[width=1.0\textwidth]{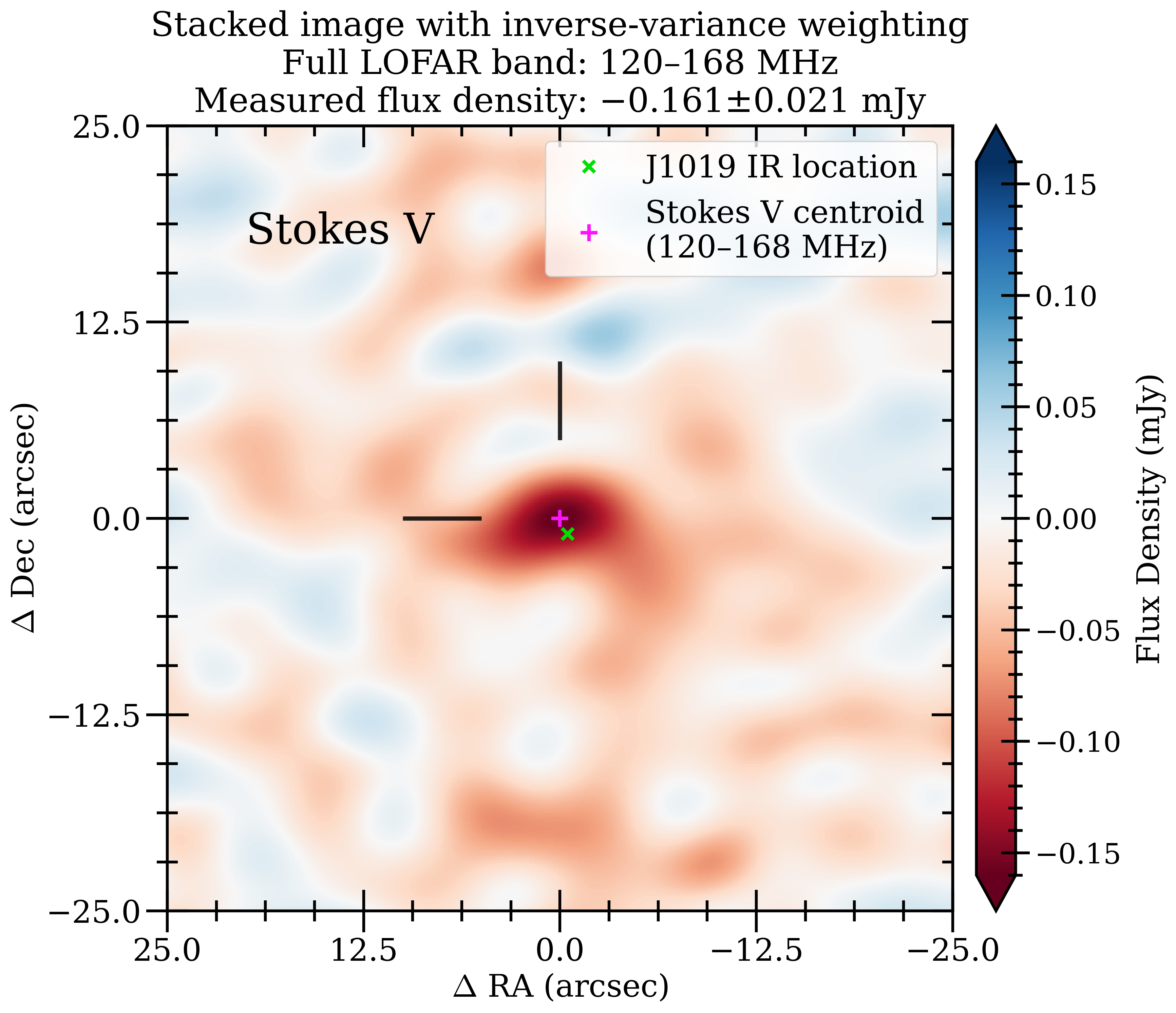}}
    \caption{Stacked Stokes V image of J1019+65 with inverse-variance weighting from all LOFAR data (2017--2023), centred at the radio centroid (marked in magenta). The flux density of the radio centroid (peak pixel) is around \qty{-0.16}{mJy}. The $1\sigma$ noise of the image is around \qty{-0.02}{mJy} The proper-motion-corrected location of J1019+65 derived from infrared observation is marked by a lime cross, approximately \qty{0.5}{\arcsec} away from the radio centroid.}
    \label{fig:stacked-image-v-all}
\end{figure}

\begin{figure}
    \centering
    \resizebox{\hsize}{!}{\includegraphics[width=1.0\textwidth]{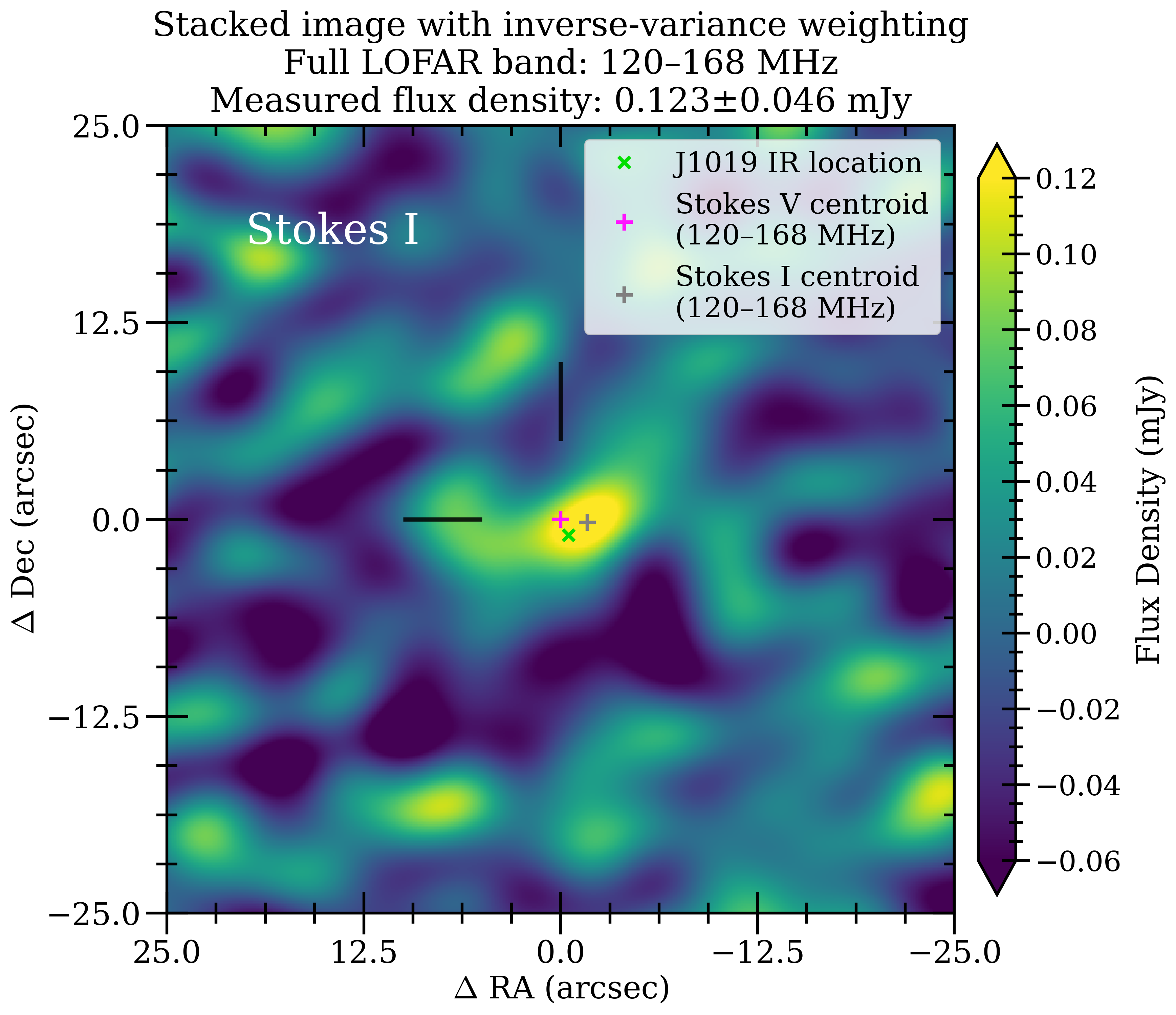}}
    \caption{Same as Fig.~\ref{fig:stacked-image-v-all}, but in Stokes I.}
    \label{fig:stacked-image-i-all}
\end{figure}

As confirmation that the emission detected in our LS analysis originates from J1019+65, Figure~\ref{fig:stacked-image-v-all} shows the inverse variation-weighted average image of J1019+65 made by stacking Stokes V images of individual LOFAR epochs from 2017 to 2023. Here we detected a source with flux density of \num{-0.161 \pm 0.021}\,mJy near the expected location of J1019+65 in the stacked image.
The proper-motion-corrected location of J1019+65 derived from infrared observations (cross marker in lime colour; \citealp{kirkpatrick2019}) is approximately \qty{0.5}{\arcsec} away from the radio centroid, which has roughly the same value as the Gaussian-equivalent standard deviation ($\sigma \approx \qty{0.5}{\arcsec}$) in the absolute radio astrometry of LOFAR \citep{lotss_dr2}. As such, we are confident of the association between J1019+65 and the radio source in the stacked image.

Comparing the stacked image in Stokes V (Fig.~\ref{fig:stacked-image-v-all}) with that in Stokes I (Fig.~\ref{fig:stacked-image-i-all}, we can see that the source is $\approx 100 \%$ circularly polarised, which is consistent with ECM emission.

\section{Results and Discussion}
\label{sec:results}
\subsection{Handedness and emission geometry}
As seen in Figures \ref{fig:phasefolded_lc_3h} and \ref{fig:phasefolded_lc_1h}, the emission of J1019+65 occurs in specific phases and has a preferred Stokes V handedness. Note that we define the sign of circularly polarised emission as left-hand circularly polarised emission minus right-hand circularly polarised emission \citep{v-lotss}. The bright burst arrives at a phase corresponding to the peak in the rotationally modulated emission.
A preference in both the emission polarisation and the rotational phase of the radio source is also observed in Jupiter \citep{marques2017}. In Jupiter, right-hand circular polarised emission (stemming from the northern hemisphere) is more frequent than its left-hand southern counterpart due to the stronger magnetic field amplitude in the northern hemisphere \citep{zarka1996,bonfond2013}. The preferential detection of Jovian radio emission at certain rotational phase is due to ECM beaming, as discussed in Sect.~\ref{sec:results-single}. Therefore, the phenomenology of rotational modulated emission in J1019+65 shares several similarities with the Jovian case, suggesting a Jupiter-like radio engine operating in J1019+65. 

The origin of the persistent emission of J1019+65 can therefore be interpreted as follows: The magnetic field lines of J1019+65 host ECM emission in a large region covering all azimuthal sectors. As the magnetosphere of J1019+65 rotates, the ECM emission beams from different azimuthal sectors come into and out of our line of sight. The rotationally modulated emission of J1019+65 detected within a particular phase is simply the collective ECM emission from all the visible beams during said time frame, and its maximum magnitude occurs when the maximum range of sectors is visible to the observer. Typically, magnetic obliquity restricts the visible period to a small fraction of the rotation period and could make the emission from both magnetic poles visible at some fraction of the rotation period \citep{williams2017}. In contrast, J1019+65 shows a persistent emission of the same handedness. This could indicate that the magnetic obliquity is smaller than the angular thickness of the emission cone, thereby keeping the source from a single magnetic pole in view for most of its rotation.

The detected radio burst of J1019+65 occurred near the peak of the phase-folded light curves as seen in Figs.~\ref{fig:phasefolded_lc_3h}~and~\ref{fig:phasefolded_lc_1h}. 
The location of the burst is also consistent with a Jupiter-like engine because we have the highest chance of seeing anomalously bright ECM-induced radio bursts when the persistent emission is detected at the highest significance, which should correspond to the rotational phase at which the maximum amount of ECM conical beams are towards our line of sight. We also note that the flux density of the burst of J1019+65 is around several mJy, which is $\sim 1$ order of magnitude brighter than its persistent emission. This is consistent with what was seen in Jupiter as well \citep{zarka_bible2007}.

\subsection{The true nature of $f_2$}
\label{sec:f2-true-nature}
Finding two distinct frequencies in the periodogram is unusual and requires an explanation. There are four possibilities for the secondary peak at $f_2$: (i) a spurious frequency caused by aliasing, (ii) some harmonic of $f_1$, (iii) a true frequency corresponding to the rotation period of the other brown dwarf, or (iv) a true frequency due to satellite interaction. Let us tackle each possibility individually in this order.

\paragraph{Aliasing:} Like any other method based on Fourier transform, the LS periodograms can have aliased power. In general, for an observation whose window function has a strong power at $\delta f$, one can expect an observed peak $f_{\rm peak}$ to be an alias if there exist peaks at $\lvert {f_{\rm peak} \pm n \delta f} \rvert$, where $n \in \mathbb{N}^+$ is small ($n \sim 1$). With that in mind, we inspected the window function for our cross-epoch analysis (plotted in \ref{fig-app:window}) and identified the different prominent peaks. For example, our ``sampling'' rate of 4 minutes corresponds to $\delta f = \qty{15}{hour^{-1}}$, so clearly this cannot be responsible for creating aliases of $f_1$ at $f_2$. We were unable to identify a peak that could explain the value of $f_2$ and $f_1$ based on aliasing. 

\paragraph{Harmonic:} If the signal is periodic but not sinusoidal, its power in the periodogram will be present at the fundamental frequency $f$ also at higher harmonics $nf$, where $n \in \{2,3,4,\ldots\}$. This is especially relevant for Jupiter-like radio-loud objects such as J1019+65, since (i) we anticipate a light curve that resembles an on-off function, and (ii) there exist certain magnetic configurations that allow a brown dwarf to pulsate more than once per rotation; Jupiter typically emits 2 pulses for every rotation. As $f_2/f_1 = 3.91$, the closest harmonic number is $n=4$. This results in a residual of $4f_1 - f_2 = 0.029 \pm 0.025 \,\unit{hour^{-1}}$. In comparison, the frequency uncertainty of $f_1$ in the original LoTSS DR2 data shown in Fig.~\ref{fig-app:lofar-2017-multi-lcls} has a value $>3$ times smaller than this residual. Moreover, there are no signs of a significant harmonic peak at $2f_1$ or $3f_1$ in the LS periodogram (recall Fig.~\ref{fig:cross-epoch-ls}), which is contrary to what was expected if $f_2$ is indeed the fourth harmonic. Furthermore, the LS power at $f_2$ can exceed that at $f_1$ in some spectral bands as seen in Fig.~\ref{fig-app:cross-epoch-ls-multi}. If $f_2$ were to be a harmonic of $f_1$, $f_2$ should not have any LS power in the 136--168\,MHz band, and yet it does. For the above reasons, $f_2$ is unlikely to be a higher harmonic of $f_1$.

\paragraph{Rotation of the second brown dwarf:} Since J1019+65 is a brown dwarf binary, $f_2$ may correspond to the rotation rate of the ``second'' brown dwarf, given that $f_1$ corresponds to the ``first'' brown dwarf. In this scenario, both brown dwarfs in J1019+65 emit persistent ECM emissions in the LOFAR band; since LOFAR does not have sufficient angular resolution, the emissions of both brown dwarfs were superposed, thus creating a light curve with two periodic signatures $P_1 = 3.07$\,h and $P_2 = 0.79$\,h.
Using the estimation of J1019+65's physical parameters by \citep{harish_j1019}, we calculated the rotation break-up period to be 0.45\,h for the more massive T5.5 brown dwarf and 0.51\,h for the lighter T7.0 brown dwarf in J1019+65, so a 0.79\,h period is physically possible.
If true, the second brown dwarf would be the fastest rotating brown dwarf ever detected (cf. \citealp{tannock2021}). However, interpreting $P_2$ as the rotation period can be fraught \citep[see for e.g. ][]{williams2017} without independent confirmation from infrared photometry. For instance, it is also plausible that the radio period is a harmonic of the true rotation period. 
Regardless, for this scenario to work, both radio emitting objects must have comparable radio luminosities, which is somewhat contrived. Follow-up infrared observations on J1019+65 may be the key to determining the nature of $f_2$; a detection of these ($P_1$ and $P_2$) periodicities in the infrared light curve of J1019+65 due to the anticipated rotational modulation would definitively confirm the authenticity of $f_2$. Moreover, using adaptive optics, the two components of J1019+65 were easily resolved \citep{harish_j1019}. Therefore, further infrared observations could also help determine which brown dwarf has the longer or shorter period.

\paragraph{Satellite interaction:} Based on the Jovian analogy, an intriguing possibility is that the two periodicities come from two emission components: one modulated by the rotation of the emitter (0.79\,h) and another modulated by the orbital period of a companion satellite (3.07\,h). Are such short rotation periods even feasible? To answer this, we first checked the Roche limit by computing the net force on a test particle at the surface of the satellite. To avoid disintegration, the centrifugal force plus the gravitational force due to the satellite's mass must exceed the gravitational force of the brown dwarf at the surface of the satellite. The calculations show that if the brown dwarf mass is $30\,{\rm M}_{\rm J}$, then disintegration of the satellite can only be prevented if the mean density of the satellite exceeds $\approx 4\,{\rm g}\,{\rm cm}^{-3}$. Such a high density is unlikely for a rocky planet, but possible for a massive gas-giant planet or another brown dwarf. Next, we considered the tidal evolution of the orbit. The putative satellite is beyond the co-rotation radius, which means its orbital distance will increase due to tidal forces. Equations for the approximate timescale for such an evolution have been derived by \citet{2011A&A...535A..94B}. We used their Eq.~2 with an eccentricity of 0, primary and secondary masses of $30\,{\rm M}_{\rm J}$ and $5\,{\rm M}_{\rm J}$ respectively, primary and secondary radii equal to the Jovian radius and an orbital period of 3\,h as characteristic values. We find that the orbital evolution timescale is under $10^6\,{\rm yr}$ even if we assume that the putative planet is perfectly tidally locked to the brown dwarf. Because we have no indication of J1019+65 being a very young system, the short tidal evolution timescale is a serious problem for the existence of a planet in such a short orbit.

\subsection{Radio-loud fraction and duty ratio of brown dwarfs}
\label{sec:FD-constraints}
J1019+65 was the only brown dwarf system within 25\,pc detected in LoTSS DR2 among hundreds of catalogued brown dwarfs and within the survey footprint. Therefore, prior to our follow-up radio observations, it was reasonable to assume that J1019+65 has some atypical property (e.g. binarity, spectral type, and rotation rate) that led to a detectable level of radio emission. Our follow-up exposures combined with the 8h-long LoTSS exposure amount to 52 hours over two years, in which we only detected one radio burst lasting for several minutes. Because J1019+65 was identified due to its bright 4-minute burst, we are led to an interesting question: Was J1019+65 discovered because it is atypical, or is J1019+65 typical and it happened to be in a high-flux-density state during the LoTSS DR2 observations? In other words, are radio-emitting brown dwarfs rare or is radio emission common among brown dwarfs, but a flux-limited surveys only picks out the small numbers of brown dwarfs that happen to be in a high-activity state?

\subsubsection{Rarity vs Activity}
\label{sec:rarity-vs-activity}
We shall first explain the ``rarity'' and ``activity'' of J1019+65 within a mathematical context: Let $F_{\rm radio} = N_{\rm radio}/N_{\rm total}$ and $\expected{D}_{\rm BD}$ be the fraction of brown dwarfs that are ``radio-loud'' and the expected (average) duty ratio of a ``radio-loud'' brown dwarf respectively. In this context, we define a ``radio-loud'' brown dwarf as one that is capable of emitting ECM emission detectable in a typical 8h LoTSS DR2 exposure, which has a median continuum noise of $95\,\mu$Jy in the $120-168$\,MHz band. For our 4h observations, this corresponds to a median continuum noise of around $134\,\mu$Jy.
indicate at what phase relative to the phase folded light curves it corresponds to
Therefore, the radio detectability of a brown dwarf mainly hinges on whether said brown dwarf can generate such ECM emissions in the first place, and if so, how often we expect the brown dwarf to generate them. $F_{\rm radio}$ encapsulates the former, while $\expected{D}_{\rm BD}$ the latter. 

Intuitively, one can see that the two quantities of $F_{\rm radio}$ and $\expected{D}_{\rm BD}$ correspond to the ``rarity'' and ``activity'' of a radio-loud brown dwarf respectively: a small value of $F_{\rm radio}$ implies that radio-loud brown dwarfs are rare, while a small value of $\expected{D}_{\rm BD}$ implies that radio-loud brown dwarfs are mostly inactive, thus leading to a small number of detections in a survey.

Mathematically, the duty ratio of a radio-loud brown dwarf is defined as $D = P_{\rm on}/P_{\rm BD}$, where $P_{\rm on}$ is the length of time for which the brown dwarf's radio signal is ``on'' in a single rotation period $P_{\rm BD} \equiv P_{\rm on} + P_{\rm off}$ of said brown dwarf. In the context of this analysis, the radio signal of J1019+65 at time $t$ can be considered ``on'' when its ECM emission exceeds a certain detection threshold during said time $t$. This definition is equivalent to the following statement: $D_{\rm J1019}$ represents the probability of detecting significant (i.e. above detection threshold) radio emission from J1019+65 at any given time (ignoring beaming for now). We shall revisit this statement in later analysis.

\subsubsection{Constraints on $\expected{D}_{\rm BD}$ and $F_{\rm radio}$}
\label{sec:DF-constraints}

With the concept of ``rarity'' and ``activity'' mathematically defined, we now proceed to explain how constraining the duty ratio of J1019+65 can ultimately lead to constraints on $F_{\rm radio}$ and $\expected{D}_{\rm BD}$.
Recall from Sect.~\ref{sec:intro} that by the virtue of an untargeted radio sky survey, the radio detection of J1019+65 in LoTSS does not suffer from selection bias that is inherent to targeted observations (where most other radio-loud methane dwarfs were found).
Moreover, as highlighted in Sect.~\ref{sec:results-cross}, the radio emission from J1019+65 matches the characteristics of the canonical Jovian periodic pulsed emission profile previously seen from brown dwarfs and ultracool dwarfs (e.g. \citealp{hallinan2007, hallinan2008, hallinan2015}), meaning J1019+65 should be a typical radio-loud brown dwarf in this regard.
Both reasons are crucial if one were to justify any estimation of $F_{\rm radio}$ and $\expected{D}_{\rm BD}$ using the statistics of J1019+65. Therefore, we assume that the magnetic field strength of J1019+65 is not atypical for the underlying brown dwarf population.

Now, because J1019+65 is at $d=23.3$\,pc, we take all known brown dwarfs within \qty{25}{pc} as our parent sample. This gives us a total of 106 brown dwarfs (spectral types L0--Y2) within the LoTSS DR2 footprint based on the volume-complete sample by \cite{best2024}. 
Therefore, this naturally leads to the following question: Given that the total number of brown dwarfs in LoTSS DR2 is $N_{\rm LoTSS} = 106$, what is the probability of detecting only 1 radio-loud brown dwarf (i.e. J1019+65)?
Even if all brown dwarfs were to be radio-loud, only a fraction would be visible from the Earth due to beaming. And since we cannot differentiate between a brown dwarf that is not radio-loud and a radio-loud brown dwarf that is not beaming towards us, the observed fraction of radio-loud brown dwarfs is therefore $F^{'}_{\rm{radio}} \equiv F_{\rm{radio}} \times F_{\rm{beam}}$, where $F_{\rm{beam}}$ is the fraction of (radio-loud) brown dwarfs that beam towards us (and thus visible to us).

\begin{figure}
    \centering
    \resizebox{\hsize}{!}{\includegraphics[width=1.0\textwidth]{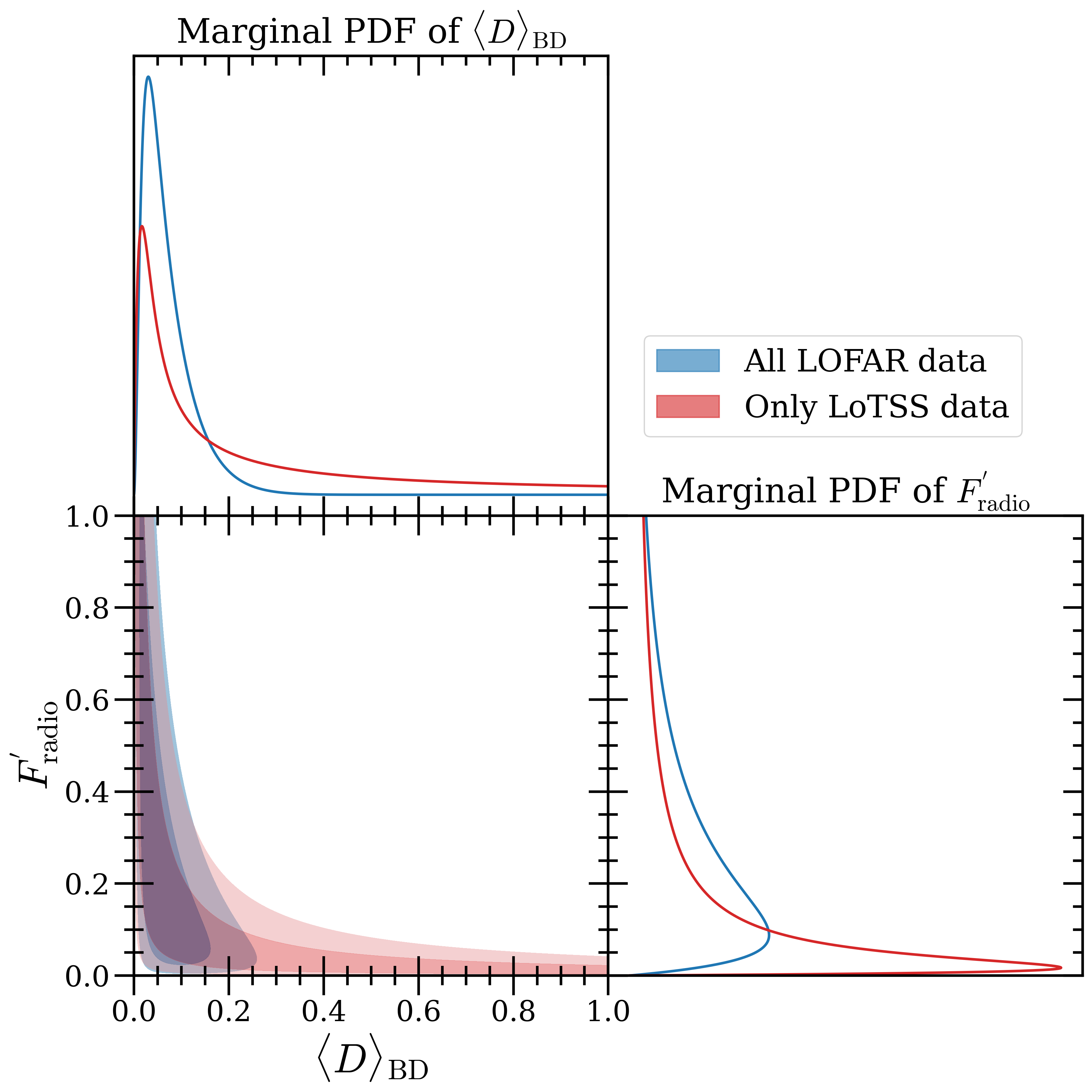}}
    \caption{Corner plot showing the covariance of the joint probability distribution function (PDF) and marginal PDFs of the two parameters: the observed fraction of radio-loud brown dwarfs $F^{'}_{\rm radio}$ in our sample and its expected duty ratio $\expected{D}_{\rm BD}$. Information plotted in red only considers the Poisson probability (Eq.~\ref{eq:poisson}) derived from LoTSS DR2 observation, whereas information in blue also considers the binomial distribution (Eq.~\ref{eq:binomial} multiplied by Eq.~\ref{eq:poisson}) derived from the follow-up LOFAR data. The 2D distribution have contours indicating the regions containing \qtylist{68;95}{\percent}.}
    \label{fig:corner-all}
\end{figure}
\begin{figure}
    \centering
    \resizebox{\hsize}{!}{\includegraphics[width=1.0\textwidth]{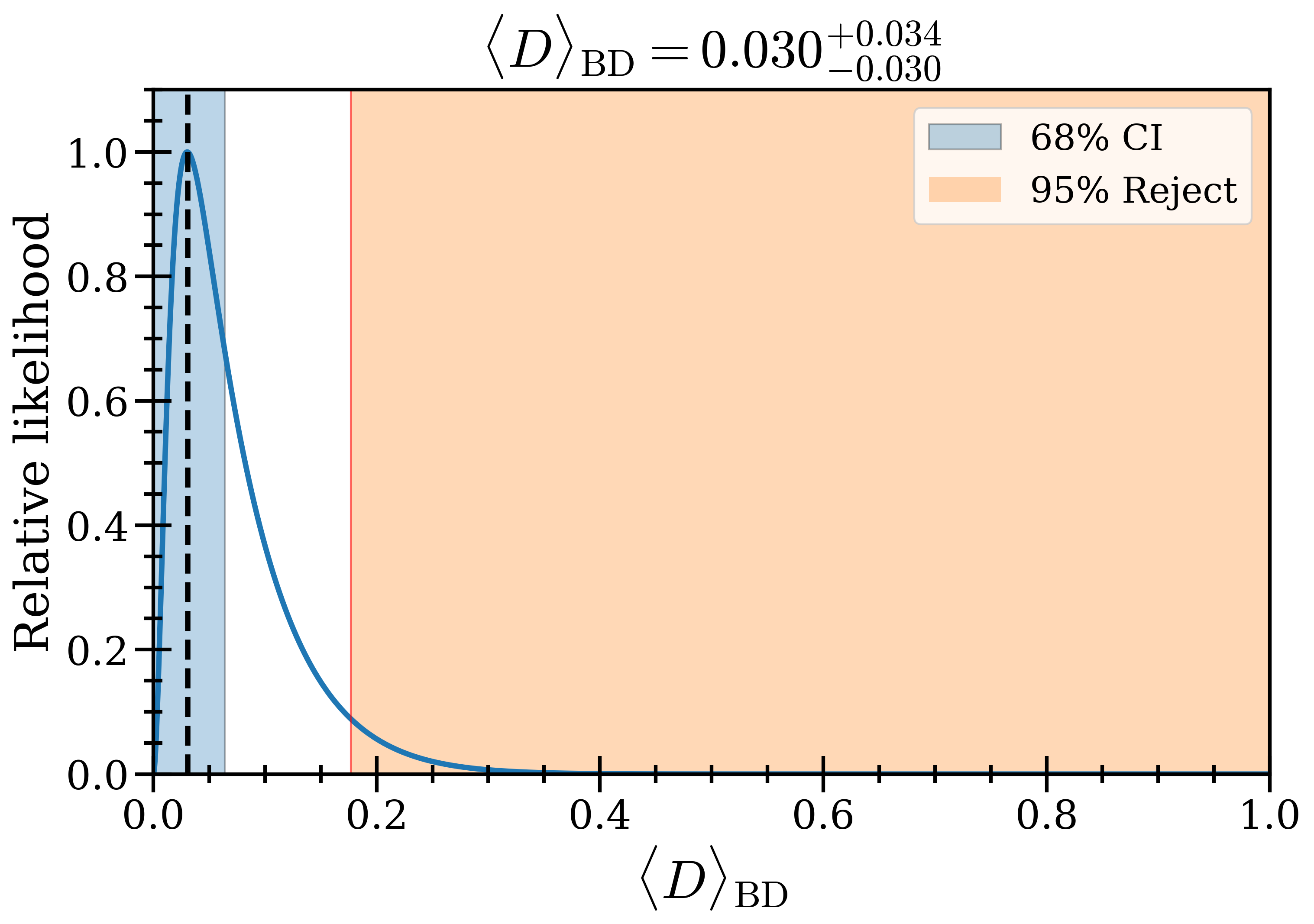}}
    \resizebox{\hsize}{!}{\includegraphics[width=1.0\textwidth]{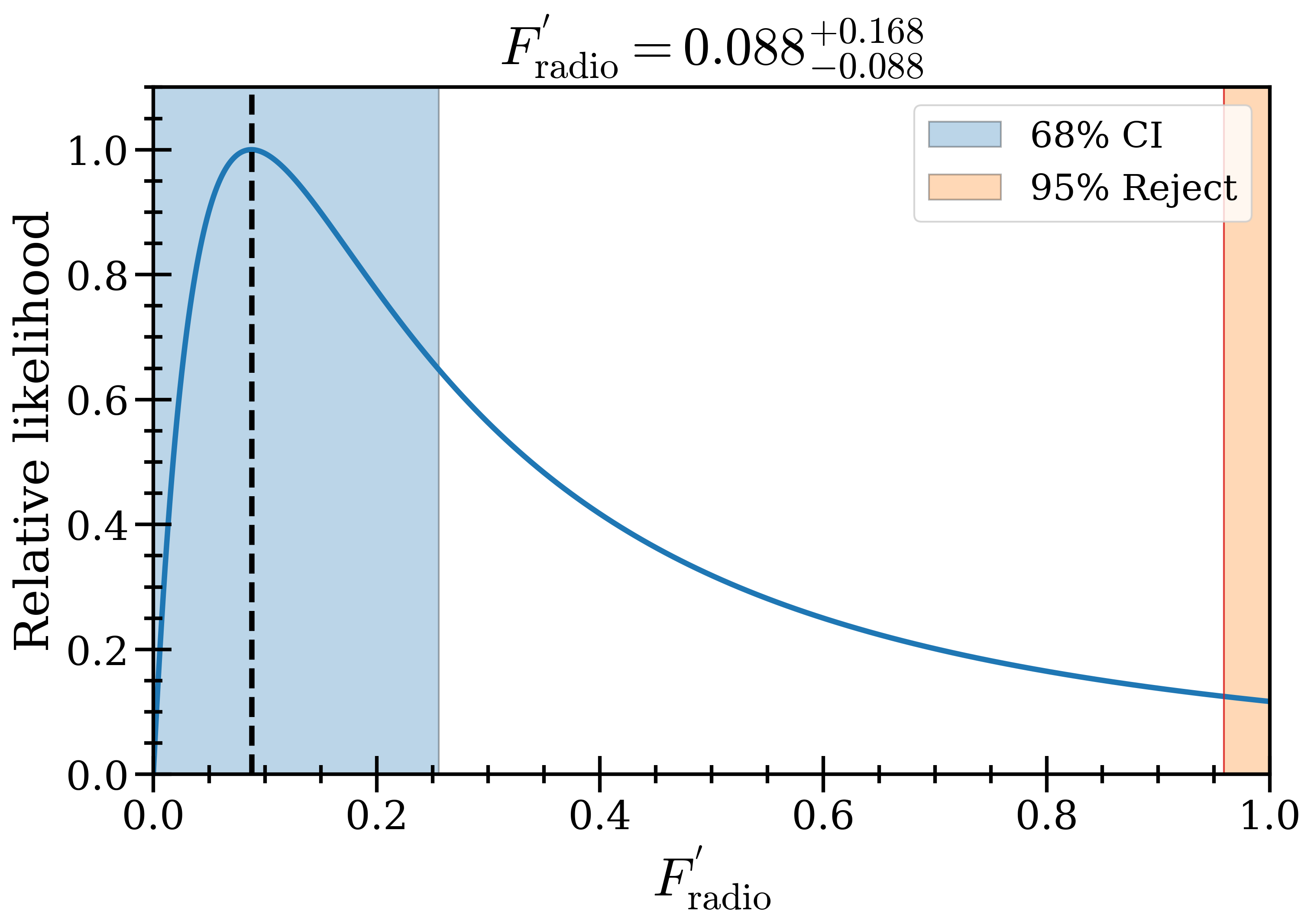}}
    \caption{Marginal probability distribution functions of $\expected{D}_{\rm BD}$ and $F^{'}_{\rm radio}$. In each plot, the relative likelihood is defined such that it is unity at the black dashed line, corresponding to the most probable value of each parameter, i.e. $\expected{D}_{\rm BD} = 0.030^{+0.034}_{-0.030}$ and $F^{'}_{\rm radio} = 0.088^{+0.168}_{-0.088}$. The error bars of the parameters correspond to \qty{68}{\percent} confidence interval (CI), as represented by the blue shaded region. The parameter value within orange shaded region can be rejected with $\geq \qty{95}{\percent}$.}
    \label{fig:marginal_DF}
\end{figure}

Finally, under these assumptions, we can proceed with determining the probability of detecting only $k=1$ brown dwarf in LoTSS DR2, which should follow a Poisson process with the rate parameter $\lambda$:
\begin{equation}
    P(\lambda,k=1) = \lambda e^{-\lambda}, \quad \lambda = N_{\rm LoTSS} \times \expected{D}_{\rm BD} \times F^{'}_{\rm radio}.
    \label{eq:poisson}
\end{equation}

Already, computing $P(\lambda)$ gives us the joint probability distribution function (PDF) of $\expected{D}_{\rm BD}$ and $F^{'}_{\rm radio}$. The singular LoTSS DR2 detection can only constrain the product of $\expected{D}_{\rm BD}$ and $F^{'}_{\rm radio}$, meaning J1019+65 could have been the singular radio-loud object, or it could be just a typical brown dwarf that happened to be in a bright state during the LoTSS DR2 exposure. This is clearly illustrated in red in Fig.~\ref{fig:corner-all}, where we computed the covariance of the posterior distribution of $\expected{D}_{\rm BD}$ and $F^{'}_{\rm radio}$ based on Eq.~\ref{eq:poisson}.

To break this degeneracy between $\expected{D}_{\rm BD}$ and $F^{'}_{\rm radio}$, we now consider the non-detections of radio pulsation in our follow-up observational campaign of J1019+65. As explained in Sect.~\ref{sec:FD-constraints}, we observed J1019+65 with LOFAR for a total of 52 hours, which corresponds to $n \approx 17$ rotations of the radio emitter in J1019+65 based on the radio-derived period $P_1 = \qty{3.0742}{hour}$. We ignore $P_2$ in this analysis\footnote{Even if $P_2$ really is the period of the radio emitter rather than $P_1$, it would have provided a tighter constraint anyway since we would have covered more than 68 rotations of the radio emitter.}. Recall that the phenomenon of radio pulsations is the consequence of rotational modulation due to beaming; intuitively, we expect to detect 1 pulse per rotation\footnote{For certain magnetic configurations, it is possible to have more than 1 pulse per rotation. For example, Jupiter typically produces 2 pulses per rotation at about \ang{40} and \ang{190} longitude.} for a brown dwarf with $D=1$. Therefore, the probability of detecting only $x=1$ bursts given that the LOFAR observations covered $n$ rotations of J1019+65 is given by:
\begin{equation}
    P(p,n=17,x=1) = \binom{n}{x}\ p^x (1-p)^{n-x} = 17p(1-p)^{16}, \quad p = D_{\rm J1019},
    \label{eq:binomial}
\end{equation} 
where $\binom{n}{x} \equiv \frac{n!}{x!(n-x)!}$ is the binomial coefficient. As explained earlier in this section, we can interpret J1019+65 as an accurate representation of a radio-loud brown dwarf in this sample owing to the virtue of untargeted survey. This is a powerful statement as it implies $D_{\rm J1019} \simeq \expected{D}_{\rm BD}$, hence we can constrain $\expected{D}_{\rm BD}$ simply by estimating the duty ratio of J1019+65. As we can see in Eq.~\ref{eq:binomial}, the binomial distribution depends solely on the duty ratio, unlike Eq.~\ref{eq:poisson}.

The likelihood of our LoTSS DR2 and follow-up observations is thus given by the product of Eq.~\ref{eq:poisson} and Eq.~\ref{eq:binomial}.
The result is shown in blue in Fig.~\ref{fig:corner-all}; compared to the data in red that only contains the LoTSS data, the extra restriction provided by Eq.~\ref{eq:binomial} greatly improve the constraints on $\expected{D}_{\rm BD}$ and consequently $F^{'}_{\rm radio}$. Figure~\ref{fig:marginal_DF} shows a detailed version of the marginal PDFs presented in Fig.~\ref{fig:corner-all}. Both curves in Fig.~\ref{fig:marginal_DF} are unimodal. Therefore, using the method of maximum likelihood estimation (MLE), we obtained $\expected{D}_{\rm BD} = 0.030^{+0.034}_{-0.030}$ and $F^{'}_{\rm radio} = 0.088^{+0.168}_{-0.088}$. Moreover, we can reject any value of $\expected{D}_{\rm BD} > 0.177$ and $F^{'}_{\rm radio} > 0.959$ with $\geq \qty{95}{\percent}$ confidence. The rejection of a large $\expected{D}_{\rm BD}$ implies that in order to capture the moment of a radio burst from a brown dwarf, one must either (i) observe a radio-loud brown dwarf for many rotations, or (ii) observe many brown dwarfs at the same time (e.g. a sky survey). Therefore, if LoTSS were to be repeated, we would expect to find new radio-loud brown dwarfs with detected burst. Furthermore, we can estimate $F_{\rm{beam}} = \Omega / 4\pi \approx \qty{12.7}{\percent}$ based on the instantaneous solid angle of the measured Jovian emission beam $\Omega = \qty{1.6}{\steradian}$ \citep{solid-angle_zarka2004}. Note that this value serves as a lower limit of $F_{\rm{beam}}$; for brown dwarfs with magnetic obliquity, their beaming pattern would thus sweep across the sky, as the magnetic pole wobbles around when observing over a whole rotation period. This estimate gives us $F_{\rm radio} = F^{'}_{\rm radio}/F_{\rm{beam}} \lesssim 0.691$, meaning that brown dwarfs such as J1019+65 cannot be too rare. In comparison, our estimation of $F_{\rm radio}$ is consistent with the work by \cite{kao2024}, which showed that the fraction of ultracool dwarf binary systems that emit persistent radio emission is $0.56^{+0.11}_{-0.11}$. Note that the focus of their work is on the incoherent (gyro)synchrotron emission stemming from radiation belts of brown dwarfs rather than coherent ECM emission.

\section{Conclusions and outlook}
\label{sec:conclusion}

Our follow-up observational campaign of J1019+65 with LOFAR led to 52 hours of observations that span around six years. We only detected a periodic signal at 120--136\,MHz in one out of ten individual LOFAR epochs using LS periodogram, while the rest of the epochs were all non-detection. Moreover, we did not detect any radio burst from J1019+65 similar to the original pulse by \cite{harish_j1019} in any of our follow-up observations. 

Despite the lack of radio bursts, we found that J1019+65 shows persistent circularly polarised emission that is periodically modulated. We find two distinct periodicities: a new periodic signature of $P_2 = 0.79$\,h and the 3h period initially identified by \citet{harish_j1019}. We find that the periodogram peak corresponding to the new period in the LS periodogram is not a spurious signal due to aliasing, harmonic structure, or the window function. If $f_2$ is indeed the rotation rate of the second brown dwarf in J1019+65, it would have the shortest rotation period of any brown dwarf to date, although an association of the two peaks to the two objects of the binary must await independent conformation, possibly from IR band photometry. Moreover, the ratio between the two inferred periods is about 4; we don't have systematic samples of T dwarf binary periods for comparison but in low-mass stellar binaries, the two components tend to have similar rotation periods (see e.g. \citealp{harding2013, stauffer2016}). If the same trend continues for cold brown dwarfs, then this period ratio would be unusual.

Based on the non-detection of bursts in most of our LOFAR observations, we computed the posterior distribution for the fraction of radio-loud brown dwarfs and the radio duty cycle of a typical brown dwarf, allowing us to constrain the rarity of a radio-loud brown dwarf and the average activity of a radio-loud brown dwarf. As such, we were able to statistically determine that radio-loud brown dwarfs in the LOFAR band should not be too rare ($F_{\rm radio} \lesssim 0.691$); we just need to stare at brown dwarfs for long enough due to their low duty ratio on average ($\expected{D} = 0.030^{+0.034}_{-0.030}$). In comparison, \cite{kao2024} showed that the fraction of ultracool dwarf binary systems that emit persistent radio emission is $0.56^{+0.11}_{-0.11}$, which is consistent with our estimation, although it should be noted that the persistent emission in their work stems from synchrotron radiation, not ECM emission.

\begin{acknowledgements}
T.W.H.Y. and H.K.V. acknowledge funding from EOSC Future (Grant Agreement no. 101017536) projects funded by the European Union's Horizon 2020 research and innovation programme. HKV acknowledges funding from the European Research Council via the starting grant `STORMCHASER' (grant number 101042416) and from the Dutch research council under the talent programme (Vidi grant VI.Vidi.203.093). JRC acknowledges funding from the European Union via the European Research Council (ERC) grant Epaphus (project number 101166008). LOFAR is the Low Frequency Array designed and constructed by ASTRON. It has observing, data processing, and data storage facilities in several countries, which are owned by various parties (each with their own funding sources), and which are collectively operated by the ILT foundation under a joint scientific policy. The ILT resources have benefited from the following recent major funding sources: CNRS-INSU, Observatoire de Paris and Universit\'e d'Orl\'eans, France; BMBF, MIWF-NRW, MPG, Germany; Science Foundation Ireland (SFI), Department of Business, Enterprise and Innovation (DBEI), Ireland; NWO, The Netherlands; The Science and Technology Facilities Council, UK; Ministry of Science and Higher Education, Poland; The Istituto Nazionale di Astrofisica (INAF), Italy.
\end{acknowledgements}

\bibliographystyle{aa}
\bibliography{ref}

\begin{thebibliography}{49}
\expandafter\ifx\csname natexlab\endcsname\relax\def\natexlab#1{#1}\fi

\bibitem[{{Astropy Collaboration} {et~al.}(2022){Astropy Collaboration}, {Price-Whelan}, {Lim}, {Earl}, {Starkman}, {Bradley}, {Shupe}, {Patil}, {Corrales}, {Brasseur}, {N{"o}the}, {Donath}, {Tollerud}, {Morris}, {Ginsburg}, {Vaher}, {Weaver}, {Tocknell}, {Jamieson}, {van Kerkwijk}, {Robitaille}, {Merry}, {Bachetti}, {G{"u}nther}, {Aldcroft}, {Alvarado-Montes}, {Archibald}, {B{'o}di}, {Bapat}, {Barentsen}, {Baz{'a}n}, {Biswas}, {Boquien}, {Burke}, {Cara}, {Cara}, {Conroy}, {Conseil}, {Craig}, {Cross}, {Cruz}, {D'Eugenio}, {Dencheva}, {Devillepoix}, {Dietrich}, {Eigenbrot}, {Erben}, {Ferreira}, {Foreman-Mackey}, {Fox}, {Freij}, {Garg}, {Geda}, {Glattly}, {Gondhalekar}, {Gordon}, {Grant}, {Greenfield}, {Groener}, {Guest}, {Gurovich}, {Handberg}, {Hart}, {Hatfield-Dodds}, {Homeier}, {Hosseinzadeh}, {Jenness}, {Jones}, {Joseph}, {Kalmbach}, {Karamehmetoglu}, {Ka{l}uszy{'n}ski}, {Kelley}, {Kern}, {Kerzendorf}, {Koch}, {Kulumani}, {Lee}, {Ly}, {Ma}, {MacBride}, {Maljaars}, {Muna}, {Murphy}, {Norman}, {O'Steen},
  {Oman}, {Pacifici}, {Pascual}, {Pascual-Granado}, {Patil}, {Perren}, {Pickering}, {Rastogi}, {Roulston}, {Ryan}, {Rykoff}, {Sabater}, {Sakurikar}, {Salgado}, {Sanghi}, {Saunders}, {Savchenko}, {Schwardt}, {Seifert-Eckert}, {Shih}, {Jain}, {Shukla}, {Sick}, {Simpson}, {Singanamalla}, {Singer}, {Singhal}, {Sinha}, {Sip{H{o}}cz}, {Spitler}, {Stansby}, {Streicher}, {{{S}}umak}, {Swinbank}, {Taranu}, {Tewary}, {Tremblay}, {Val-Borro}, {Van Kooten}, {Vasovi{'c}}, {Verma}, {de Miranda Cardoso}, {Williams}, {Wilson}, {Winkel}, {Wood-Vasey}, {Xue}, {Yoachim}, {Zhang}, {Zonca}, \& {Astropy Project Contributors}}]{astropy:2022}
{Astropy Collaboration}, {Price-Whelan}, A.~M., {Lim}, P.~L., {et~al.} 2022, \apj, 935, 167

\bibitem[{{Astropy Collaboration} {et~al.}(2018){Astropy Collaboration}, {Price-Whelan}, {Sip{\H{o}}cz}, {G{\"u}nther}, {Lim}, {Crawford}, {Conseil}, {Shupe}, {Craig}, {Dencheva}, {Ginsburg}, {Vand erPlas}, {Bradley}, {P{\'e}rez-Su{\'a}rez}, {de Val-Borro}, {Aldcroft}, {Cruz}, {Robitaille}, {Tollerud}, {Ardelean}, {Babej}, {Bach}, {Bachetti}, {Bakanov}, {Bamford}, {Barentsen}, {Barmby}, {Baumbach}, {Berry}, {Biscani}, {Boquien}, {Bostroem}, {Bouma}, {Brammer}, {Bray}, {Breytenbach}, {Buddelmeijer}, {Burke}, {Calderone}, {Cano Rodr{\'\i}guez}, {Cara}, {Cardoso}, {Cheedella}, {Copin}, {Corrales}, {Crichton}, {D'Avella}, {Deil}, {Depagne}, {Dietrich}, {Donath}, {Droettboom}, {Earl}, {Erben}, {Fabbro}, {Ferreira}, {Finethy}, {Fox}, {Garrison}, {Gibbons}, {Goldstein}, {Gommers}, {Greco}, {Greenfield}, {Groener}, {Grollier}, {Hagen}, {Hirst}, {Homeier}, {Horton}, {Hosseinzadeh}, {Hu}, {Hunkeler}, {Ivezi{\'c}}, {Jain}, {Jenness}, {Kanarek}, {Kendrew}, {Kern}, {Kerzendorf}, {Khvalko}, {King}, {Kirkby}, {Kulkarni},
  {Kumar}, {Lee}, {Lenz}, {Littlefair}, {Ma}, {Macleod}, {Mastropietro}, {McCully}, {Montagnac}, {Morris}, {Mueller}, {Mumford}, {Muna}, {Murphy}, {Nelson}, {Nguyen}, {Ninan}, {N{\"o}the}, {Ogaz}, {Oh}, {Parejko}, {Parley}, {Pascual}, {Patil}, {Patil}, {Plunkett}, {Prochaska}, {Rastogi}, {Reddy Janga}, {Sabater}, {Sakurikar}, {Seifert}, {Sherbert}, {Sherwood-Taylor}, {Shih}, {Sick}, {Silbiger}, {Singanamalla}, {Singer}, {Sladen}, {Sooley}, {Sornarajah}, {Streicher}, {Teuben}, {Thomas}, {Tremblay}, {Turner}, {Terr{\'o}n}, {van Kerkwijk}, {de la Vega}, {Watkins}, {Weaver}, {Whitmore}, {Woillez}, {Zabalza}, \& {Astropy Contributors}}]{astropy:2018}
{Astropy Collaboration}, {Price-Whelan}, A.~M., {Sip{\H{o}}cz}, B.~M., {et~al.} 2018, \aj, 156, 123

\bibitem[{{Astropy Collaboration} {et~al.}(2013){Astropy Collaboration}, {Robitaille}, {Tollerud}, {Greenfield}, {Droettboom}, {Bray}, {Aldcroft}, {Davis}, {Ginsburg}, {Price-Whelan}, {Kerzendorf}, {Conley}, {Crighton}, {Barbary}, {Muna}, {Ferguson}, {Grollier}, {Parikh}, {Nair}, {Unther}, {Deil}, {Woillez}, {Conseil}, {Kramer}, {Turner}, {Singer}, {Fox}, {Weaver}, {Zabalza}, {Edwards}, {Azalee Bostroem}, {Burke}, {Casey}, {Crawford}, {Dencheva}, {Ely}, {Jenness}, {Labrie}, {Lim}, {Pierfederici}, {Pontzen}, {Ptak}, {Refsdal}, {Servillat}, \& {Streicher}}]{astropy:2013}
{Astropy Collaboration}, {Robitaille}, T.~P., {Tollerud}, E.~J., {et~al.} 2013, \aap, 558, A33

\bibitem[{{Baluev}(2008)}]{baluev2008}
{Baluev}, R.~V. 2008, \mnras, 385, 1279

\bibitem[{{Berger} {et~al.}(2009){Berger}, {Rutledge}, {Phan-Bao}, {Basri}, {Giampapa}, {Gizis}, {Liebert}, {Mart{\'\i}n}, \& {Fleming}}]{berger2009}
{Berger}, E., {Rutledge}, R.~E., {Phan-Bao}, N., {et~al.} 2009, \apj, 695, 310

\bibitem[{{Best} {et~al.}(2024){Best}, {Sanghi}, {Liu}, {Magnier}, \& {Dupuy}}]{best2024}
{Best}, W. M.~J., {Sanghi}, A., {Liu}, M.~C., {Magnier}, E.~A., \& {Dupuy}, T.~J. 2024, \apj, 967, 115

\bibitem[{{Bolmont} {et~al.}(2011){Bolmont}, {Raymond}, \& {Leconte}}]{2011A&A...535A..94B}
{Bolmont}, E., {Raymond}, S.~N., \& {Leconte}, J. 2011, \aap, 535, A94

\bibitem[{{Bonfond} {et~al.}(2013){Bonfond}, {Hess}, {G{\'e}rard}, {Grodent}, {Radioti}, {Chantry}, {Saur}, {Jacobsen}, \& {Clarke}}]{bonfond2013}
{Bonfond}, B., {Hess}, S., {G{\'e}rard}, J.~C., {et~al.} 2013, \planss, 88, 64

\bibitem[{{Briggs}(1995)}]{briggs1995}
{Briggs}, D.~S. 1995, in American Astronomical Society Meeting Abstracts, Vol. 187, American Astronomical Society Meeting Abstracts, 112.02

\bibitem[{{Callingham} {et~al.}(2024){Callingham}, {Pope}, {Kavanagh}, {Bellotti}, {Daley-Yates}, {Damasso}, {Grie{\ss}meier}, {G{\"u}del}, {G{\"u}nther}, {Kao}, {Klein}, {Mahadevan}, {Morin}, {Nichols}, {Osten}, {P{\'e}rez-Torres}, {Pineda}, {Rigney}, {Saur}, {Stef{\'a}nsson}, {Turner}, {Vedantham}, {Vidotto}, {Villadsen}, \& {Zarka}}]{joe2024}
{Callingham}, J.~R., {Pope}, B.~J.~S., {Kavanagh}, R.~D., {et~al.} 2024, Nature Astronomy, 8, 1359

\bibitem[{{Callingham} {et~al.}(2023){Callingham}, {Shimwell}, {Vedantham}, {Bassa}, {O'Sullivan}, {Yiu}, {Bloot}, {Best}, {Hardcastle}, {Haverkorn}, {Kavanagh}, {Lamy}, {Pope}, {R{\"o}ttgering}, {Schwarz}, {Tasse}, {van Weeren}, {White}, {Zarka}, {Bomans}, {Bonafede}, {Bonato}, {Botteon}, {Bruggen}, {Chy{\.z}y}, {Drabent}, {Emig}, {Gloudemans}, {G{\"u}rkan}, {Hajduk}, {Hoang}, {Hoeft}, {Iacobelli}, {Kadler}, {Kunert-Bajraszewska}, {Mingo}, {Morabito}, {Nair}, {P{\'e}rez-Torres}, {Ray}, {Riseley}, {Rowlinson}, {Shulevski}, {Sweijen}, {Timmerman}, {Vaccari}, \& {Zheng}}]{v-lotss}
{Callingham}, J.~R., {Shimwell}, T.~W., {Vedantham}, H.~K., {et~al.} 2023, \aap, 670, A124

\bibitem[{{Carbary} \& {Mitchell}(2013)}]{carbary2013}
{Carbary}, J.~F. \& {Mitchell}, D.~G. 2013, Reviews of Geophysics, 51, 1

\bibitem[{{de Gasperin} {et~al.}(2019){de Gasperin}, {Dijkema}, {Drabent}, {Mevius}, {Rafferty}, {van Weeren}, {Br{\"u}ggen}, {Callingham}, {Emig}, {Heald}, {Intema}, {Morabito}, {Offringa}, {Oonk}, {Orr{\`u}}, {R{\"o}ttgering}, {Sabater}, {Shimwell}, {Shulevski}, \& {Williams}}]{linc_degasperin2019}
{de Gasperin}, F., {Dijkema}, T.~J., {Drabent}, A., {et~al.} 2019, \aap, 622, A5

\bibitem[{{Hallinan} {et~al.}(2006){Hallinan}, {Antonova}, {Doyle}, {Bourke}, {Brisken}, \& {Golden}}]{hallinan2006}
{Hallinan}, G., {Antonova}, A., {Doyle}, J.~G., {et~al.} 2006, \apj, 653, 690

\bibitem[{{Hallinan} {et~al.}(2008){Hallinan}, {Antonova}, {Doyle}, {Bourke}, {Lane}, \& {Golden}}]{hallinan2008}
{Hallinan}, G., {Antonova}, A., {Doyle}, J.~G., {et~al.} 2008, \apj, 684, 644

\bibitem[{{Hallinan} {et~al.}(2007){Hallinan}, {Bourke}, {Lane}, {Antonova}, {Zavala}, {Brisken}, {Boyle}, {Vrba}, {Doyle}, \& {Golden}}]{hallinan2007}
{Hallinan}, G., {Bourke}, S., {Lane}, C., {et~al.} 2007, \apjl, 663, L25

\bibitem[{{Hallinan} {et~al.}(2015){Hallinan}, {Littlefair}, {Cotter}, {Bourke}, {Harding}, {Pineda}, {Butler}, {Golden}, {Basri}, {Doyle}, {Kao}, {Berdyugina}, {Kuznetsov}, {Rupen}, \& {Antonova}}]{hallinan2015}
{Hallinan}, G., {Littlefair}, S.~P., {Cotter}, G., {et~al.} 2015, \nat, 523, 568

\bibitem[{{Harding} {et~al.}(2013){Harding}, {Hallinan}, {Konopacky}, {Kratter}, {Boyle}, {Butler}, \& {Golden}}]{harding2013}
{Harding}, L.~K., {Hallinan}, G., {Konopacky}, Q.~M., {et~al.} 2013, \aap, 554, A113

\bibitem[{{Higgins} {et~al.}(1997){Higgins}, {Carr}, {Reyes}, {Greenman}, \& {Lebo}}]{higgins1997}
{Higgins}, C.~A., {Carr}, T.~D., {Reyes}, F., {Greenman}, W.~B., \& {Lebo}, G.~R. 1997, \jgr, 102, 22033

\bibitem[{{Kao} {et~al.}(2016){Kao}, {Hallinan}, {Pineda}, {Escala}, {Burgasser}, {Bourke}, \& {Stevenson}}]{kao2016}
{Kao}, M.~M., {Hallinan}, G., {Pineda}, J.~S., {et~al.} 2016, \apj, 818, 24

\bibitem[{{Kao} {et~al.}(2018){Kao}, {Hallinan}, {Pineda}, {Stevenson}, \& {Burgasser}}]{kao2018}
{Kao}, M.~M., {Hallinan}, G., {Pineda}, J.~S., {Stevenson}, D., \& {Burgasser}, A. 2018, \apjs, 237, 25

\bibitem[{{Kao} \& {Pineda}(2024)}]{kao2024}
{Kao}, M.~M. \& {Pineda}, J.~S. 2024, \mnras [\eprint[arXiv]{2403.08860}]

\bibitem[{{Kipping}(2009)}]{kipping2009}
{Kipping}, D.~M. 2009, \mnras, 392, 181

\bibitem[{{Kipping} {et~al.}(2012){Kipping}, {Bakos}, {Buchhave}, {Nesvorn{\'y}}, \& {Schmitt}}]{kipping2012}
{Kipping}, D.~M., {Bakos}, G.~{\'A}., {Buchhave}, L., {Nesvorn{\'y}}, D., \& {Schmitt}, A. 2012, \apj, 750, 115

\bibitem[{{Kirkpatrick} {et~al.}(2019){Kirkpatrick}, {Martin}, {Smart}, {Cayago}, {Beichman}, {Marocco}, {Gelino}, {Faherty}, {Cushing}, {Schneider}, {Mace}, {Tinney}, {Wright}, {Lowrance}, {Ingalls}, {Vrba}, {Munn}, {Dahm}, \& {McLean}}]{kirkpatrick2019}
{Kirkpatrick}, J.~D., {Martin}, E.~C., {Smart}, R.~L., {et~al.} 2019, \apjs, 240, 19

\bibitem[{{Lomb}(1976)}]{lomb1976}
{Lomb}, N.~R. 1976, \apss, 39, 447

\bibitem[{{Marques} {et~al.}(2017){Marques}, {Zarka}, {Echer}, {Ryabov}, {Alves}, {Denis}, \& {Coffre}}]{marques2017}
{Marques}, M.~S., {Zarka}, P., {Echer}, E., {et~al.} 2017, \aap, 604, A17

\bibitem[{{Offringa} {et~al.}(2014){Offringa}, {McKinley}, {Hurley-Walker}, {Briggs}, {Wayth}, {Kaplan}, {Bell}, {Feng}, {Neben}, {Hughes}, {Rhee}, {Murphy}, {Bhat}, {Bernardi}, {Bowman}, {Cappallo}, {Corey}, {Deshpande}, {Emrich}, {Ewall-Wice}, {Gaensler}, {Goeke}, {Greenhill}, {Hazelton}, {Hindson}, {Johnston-Hollitt}, {Jacobs}, {Kasper}, {Kratzenberg}, {Lenc}, {Lonsdale}, {Lynch}, {McWhirter}, {Mitchell}, {Morales}, {Morgan}, {Kudryavtseva}, {Oberoi}, {Ord}, {Pindor}, {Procopio}, {Prabu}, {Riding}, {Roshi}, {Shankar}, {Srivani}, {Subrahmanyan}, {Tingay}, {Waterson}, {Webster}, {Whitney}, {Williams}, \& {Williams}}]{wsclean2014}
{Offringa}, A.~R., {McKinley}, B., {Hurley-Walker}, N., {et~al.} 2014, \mnras, 444, 606

\bibitem[{{Rose} {et~al.}(2023){Rose}, {Pritchard}, {Murphy}, {Caleb}, {Dobie}, {Driessen}, {Duchesne}, {Kaplan}, {Lenc}, \& {Wang}}]{rose2023}
{Rose}, K., {Pritchard}, J., {Murphy}, T., {et~al.} 2023, \apjl, 951, L43

\bibitem[{{Route} \& {Wolszczan}(2012)}]{route2012}
{Route}, M. \& {Wolszczan}, A. 2012, \apjl, 747, L22

\bibitem[{{Route} \& {Wolszczan}(2016)}]{route2016}
{Route}, M. \& {Wolszczan}, A. 2016, \apjl, 821, L21

\bibitem[{{Scargle}(1982)}]{scargle1982}
{Scargle}, J.~D. 1982, \apj, 263, 835

\bibitem[{{Shimwell} {et~al.}(2022){Shimwell}, {Hardcastle}, {Tasse}, {Best}, {R{\"o}ttgering}, {Williams}, {Botteon}, {Drabent}, {Mechev}, {Shulevski}, {van Weeren}, {Bester}, {Br{\"u}ggen}, {Brunetti}, {Callingham}, {Chy{\.z}y}, {Conway}, {Dijkema}, {Duncan}, {de Gasperin}, {Hale}, {Haverkorn}, {Hugo}, {Jackson}, {Mevius}, {Miley}, {Morabito}, {Morganti}, {Offringa}, {Oonk}, {Rafferty}, {Sabater}, {Smith}, {Schwarz}, {Smirnov}, {O'Sullivan}, {Vedantham}, {White}, {Albert}, {Alegre}, {Asabere}, {Bacon}, {Bonafede}, {Bonnassieux}, {Brienza}, {Bilicki}, {Bonato}, {Calistro Rivera}, {Cassano}, {Cochrane}, {Croston}, {Cuciti}, {Dallacasa}, {Danezi}, {Dettmar}, {Di Gennaro}, {Edler}, {En{\ss}lin}, {Emig}, {Franzen}, {Garc{\'\i}a-Vergara}, {Grange}, {G{\"u}rkan}, {Hajduk}, {Heald}, {Heesen}, {Hoang}, {Hoeft}, {Horellou}, {Iacobelli}, {Jamrozy}, {Jeli{\'c}}, {Kondapally}, {Kukreti}, {Kunert-Bajraszewska}, {Magliocchetti}, {Mahatma}, {Ma{\l}ek}, {Mandal}, {Massaro}, {Meyer-Zhao}, {Mingo}, {Mostert}, {Nair},
  {Nakoneczny}, {Nikiel-Wroczy{\'n}ski}, {Orr{\'u}}, {Pajdosz-{\'S}mierciak}, {Pasini}, {Prandoni}, {van Piggelen}, {Rajpurohit}, {Retana-Montenegro}, {Riseley}, {Rowlinson}, {Saxena}, {Schrijvers}, {Sweijen}, {Siewert}, {Timmerman}, {Vaccari}, {Vink}, {West}, {Wo{\l}owska}, {Zhang}, \& {Zheng}}]{lotss_dr2}
{Shimwell}, T.~W., {Hardcastle}, M.~J., {Tasse}, C., {et~al.} 2022, \aap, 659, A1

\bibitem[{{Shimwell} {et~al.}(2017){Shimwell}, {R{\"o}ttgering}, {Best}, {Williams}, {Dijkema}, {de Gasperin}, {Hardcastle}, {Heald}, {Hoang}, {Horneffer}, {Intema}, {Mahony}, {Mandal}, {Mechev}, {Morabito}, {Oonk}, {Rafferty}, {Retana-Montenegro}, {Sabater}, {Tasse}, {van Weeren}, {Br{\"u}ggen}, {Brunetti}, {Chy{\.z}y}, {Conway}, {Haverkorn}, {Jackson}, {Jarvis}, {McKean}, {Miley}, {Morganti}, {White}, {Wise}, {van Bemmel}, {Beck}, {Brienza}, {Bonafede}, {Calistro Rivera}, {Cassano}, {Clarke}, {Cseh}, {Deller}, {Drabent}, {van Driel}, {Engels}, {Falcke}, {Ferrari}, {Fr{\"o}hlich}, {Garrett}, {Harwood}, {Heesen}, {Hoeft}, {Horellou}, {Israel}, {Kapi{\'n}ska}, {Kunert-Bajraszewska}, {McKay}, {Mohan}, {Orr{\'u}}, {Pizzo}, {Prandoni}, {Schwarz}, {Shulevski}, {Sipior}, {Smith}, {Sridhar}, {Steinmetz}, {Stroe}, {Varenius}, {van der Werf}, {Zensus}, \& {Zwart}}]{lotss}
{Shimwell}, T.~W., {R{\"o}ttgering}, H.~J.~A., {Best}, P.~N., {et~al.} 2017, \aap, 598, A104

\bibitem[{{Stauffer} {et~al.}(2016){Stauffer}, {Rebull}, {Bouvier}, {Hillenbrand}, {Collier-Cameron}, {Pinsonneault}, {Aigrain}, {Barrado}, {Bouy}, {Ciardi}, {Cody}, {David}, {Micela}, {Soderblom}, {Somers}, {Stassun}, {Valenti}, \& {Vrba}}]{stauffer2016}
{Stauffer}, J., {Rebull}, L., {Bouvier}, J., {et~al.} 2016, \aj, 152, 115

\bibitem[{{Tannock} {et~al.}(2021){Tannock}, {Metchev}, {Heinze}, {Miles-P{\'a}ez}, {Gagn{\'e}}, {Burgasser}, {Marley}, {Apai}, {Su{\'a}rez}, \& {Plavchan}}]{tannock2021}
{Tannock}, M.~E., {Metchev}, S., {Heinze}, A., {et~al.} 2021, \aj, 161, 224

\bibitem[{{Tasse} {et~al.}(2021){Tasse}, {Shimwell}, {Hardcastle}, {O'Sullivan}, {van Weeren}, {Best}, {Bester}, {Hugo}, {Smirnov}, {Sabater}, {Calistro-Rivera}, {de Gasperin}, {Morabito}, {R{\"o}ttgering}, {Williams}, {Bonato}, {Bondi}, {Botteon}, {Br{\"u}ggen}, {Brunetti}, {Chy{\.z}y}, {Garrett}, {G{\"u}rkan}, {Jarvis}, {Kondapally}, {Mandal}, {Prandoni}, {Repetti}, {Retana-Montenegro}, {Schwarz}, {Shulevski}, \& {Wiaux}}]{tasse2021}
{Tasse}, C., {Shimwell}, T., {Hardcastle}, M.~J., {et~al.} 2021, \aap, 648, A1

\bibitem[{{Treumann}(2006)}]{treumann2006}
{Treumann}, R.~A. 2006, \aapr, 13, 229

\bibitem[{{van Weeren} {et~al.}(2021){van Weeren}, {Shimwell}, {Botteon}, {Brunetti}, {Br{\"u}ggen}, {Boxelaar}, {Cassano}, {Di Gennaro}, {Andrade-Santos}, {Bonnassieux}, {Bonafede}, {Cuciti}, {Dallacasa}, {de Gasperin}, {Gastaldello}, {Hardcastle}, {Hoeft}, {Kraft}, {Mandal}, {Rossetti}, {R{\"o}ttgering}, {Tasse}, \& {Wilber}}]{vanweeren2021}
{van Weeren}, R.~J., {Shimwell}, T.~W., {Botteon}, A., {et~al.} 2021, \aap, 651, A115

\bibitem[{{VanderPlas}(2018)}]{vanderplas2018}
{VanderPlas}, J.~T. 2018, \apjs, 236, 16

\bibitem[{{Vedantham} {et~al.}(2020){Vedantham}, {Callingham}, {Shimwell}, {Dupuy}, {Best}, {Liu}, {Zhang}, {De}, {Lamy}, {Zarka}, {R{\"o}ttgering}, \& {Shulevski}}]{harish_elegast2020}
{Vedantham}, H.~K., {Callingham}, J.~R., {Shimwell}, T.~W., {et~al.} 2020, \apjl, 903, L33

\bibitem[{{Vedantham} {et~al.}(2023){Vedantham}, {Dupuy}, {Evans}, {Sanghi}, {Callingham}, {Shimwell}, {Best}, {Liu}, \& {Zarka}}]{harish_j1019}
{Vedantham}, H.~K., {Dupuy}, T.~J., {Evans}, E.~L., {et~al.} 2023, \aap, 675, L6

\bibitem[{{Williams} \& {Berger}(2015)}]{williams2015}
{Williams}, P.~K.~G. \& {Berger}, E. 2015, \apj, 808, 189

\bibitem[{{Williams} {et~al.}(2017){Williams}, {Gizis}, \& {Berger}}]{williams2017}
{Williams}, P.~K.~G., {Gizis}, J.~E., \& {Berger}, E. 2017, \apj, 834, 117

\bibitem[{{Wu} \& {Lee}(1979)}]{wu1979}
{Wu}, C.~S. \& {Lee}, L.~C. 1979, \apj, 230, 621

\bibitem[{{Zarka}(1998)}]{zarka1998}
{Zarka}, P. 1998, \jgr, 103, 20159

\bibitem[{{Zarka}(2007)}]{zarka_bible2007}
{Zarka}, P. 2007, \planss, 55, 598

\bibitem[{{Zarka} {et~al.}(2004){Zarka}, {Cecconi}, \& {Kurth}}]{solid-angle_zarka2004}
{Zarka}, P., {Cecconi}, B., \& {Kurth}, W.~S. 2004, Journal of Geophysical Research (Space Physics), 109, A09S15

\bibitem[{{Zarka} {et~al.}(1996){Zarka}, {Farges}, {Ryabov}, {Abada-Simon}, \& {Denis}}]{zarka1996}
{Zarka}, P., {Farges}, T., {Ryabov}, B.~P., {Abada-Simon}, M., \& {Denis}, L. 1996, \grl, 23, 125

\end{thebibliography}

\appendix

\section{Supplementary figures}
\label{app:figures}

\begin{figure}
    \centering
    \resizebox{\hsize}{!}{\includegraphics[width=0.95\textwidth]{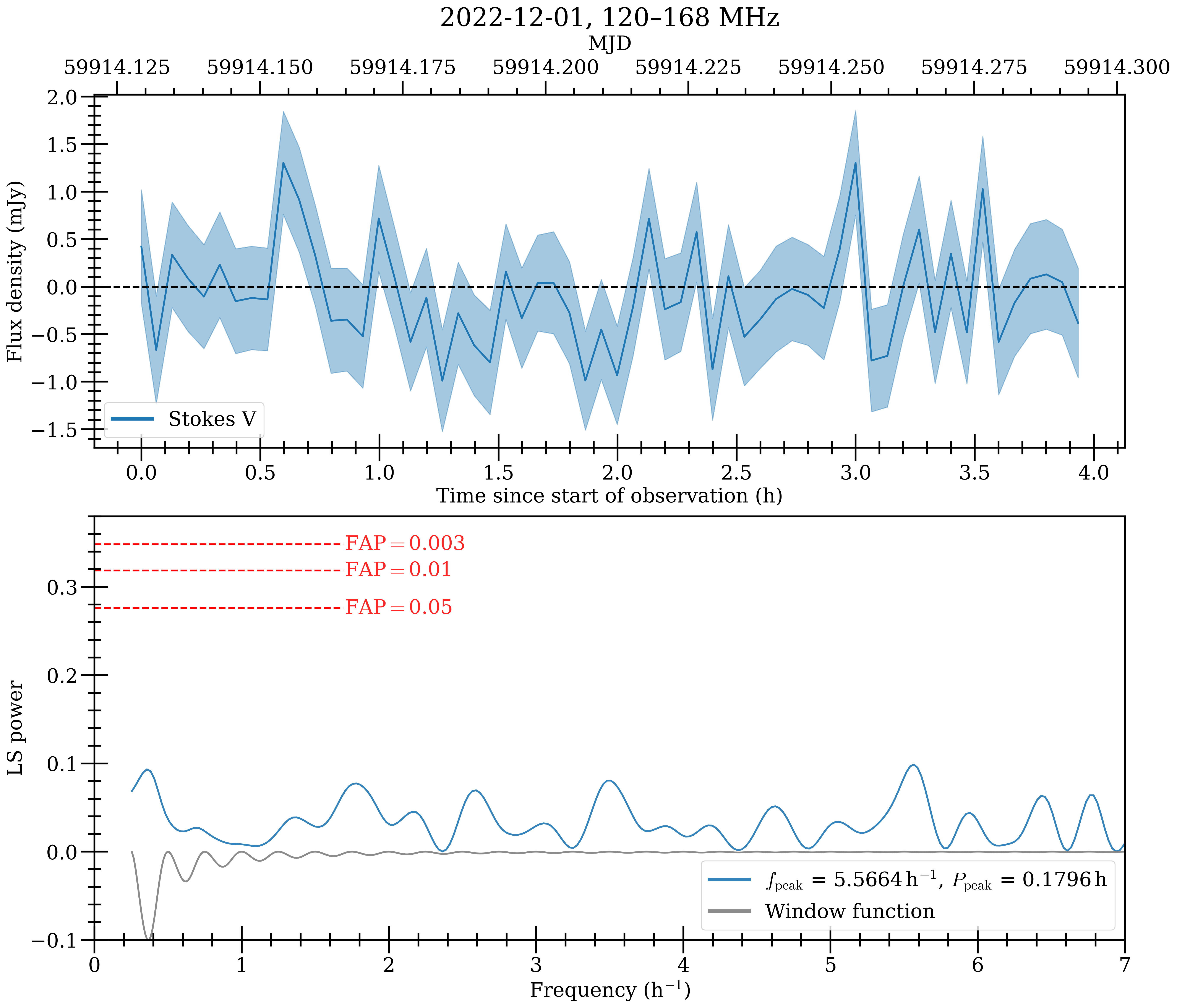}}
    \resizebox{\hsize}{!}{\includegraphics[width=0.95\textwidth]{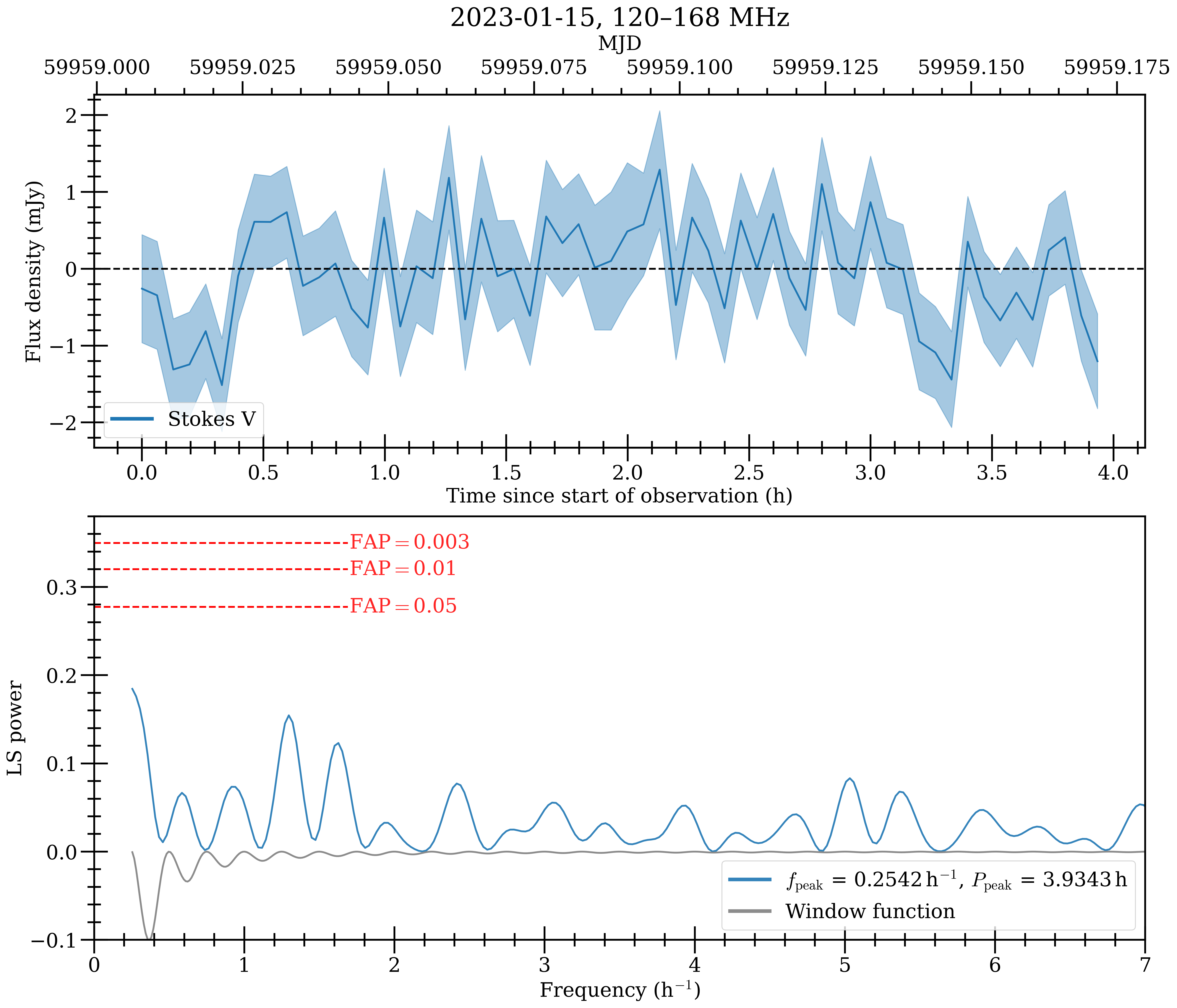}}
    \resizebox{\hsize}{!}{\includegraphics[width=0.95\textwidth]{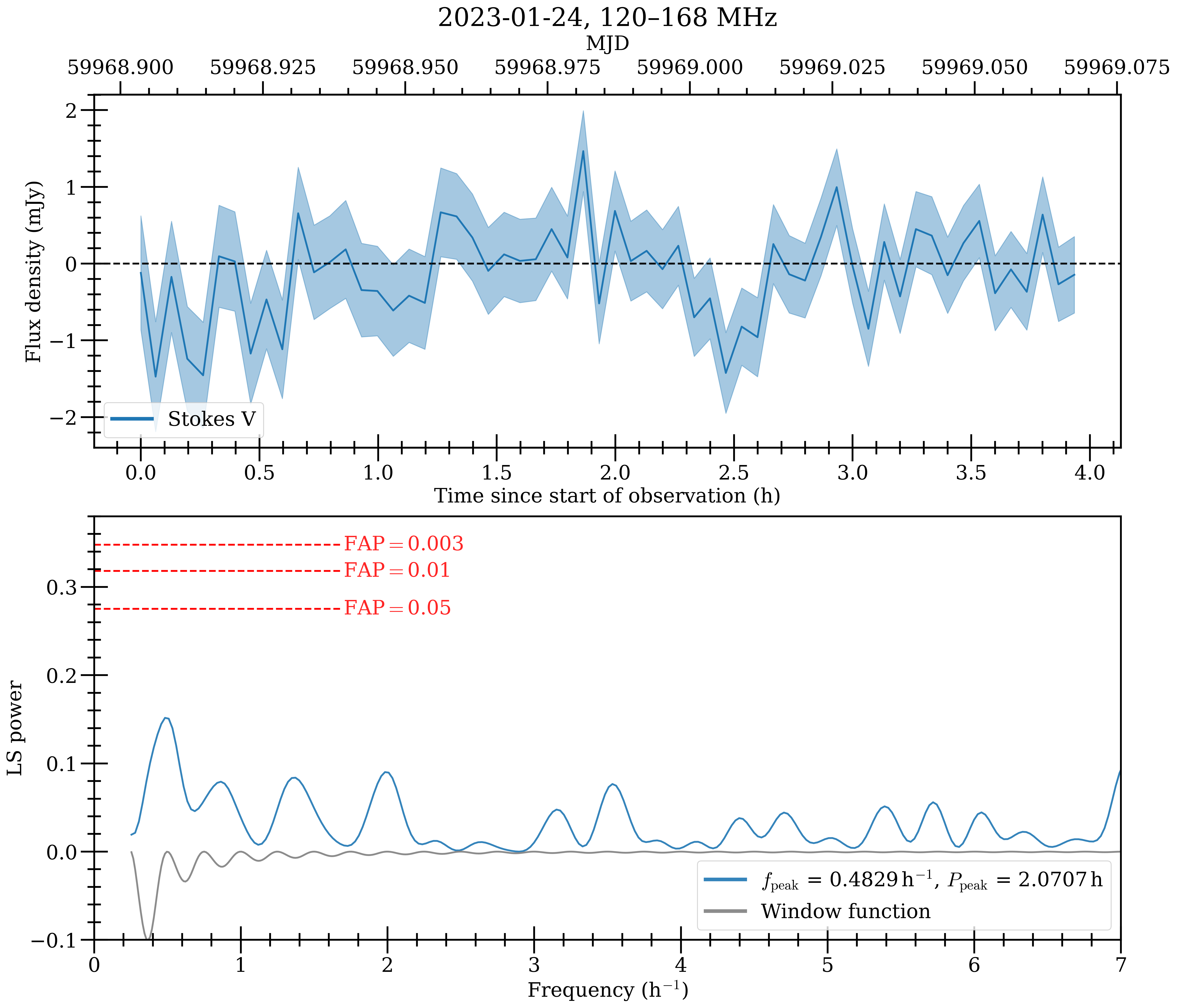}}
    \caption{Follow-up LOFAR observations of J1019+65 carried out on the first 3 dates.}
    \label{fig-app:lofar-3-obs}
\end{figure}
\begin{figure}
    \resizebox{\hsize}{!}{\includegraphics[width=0.95\textwidth]{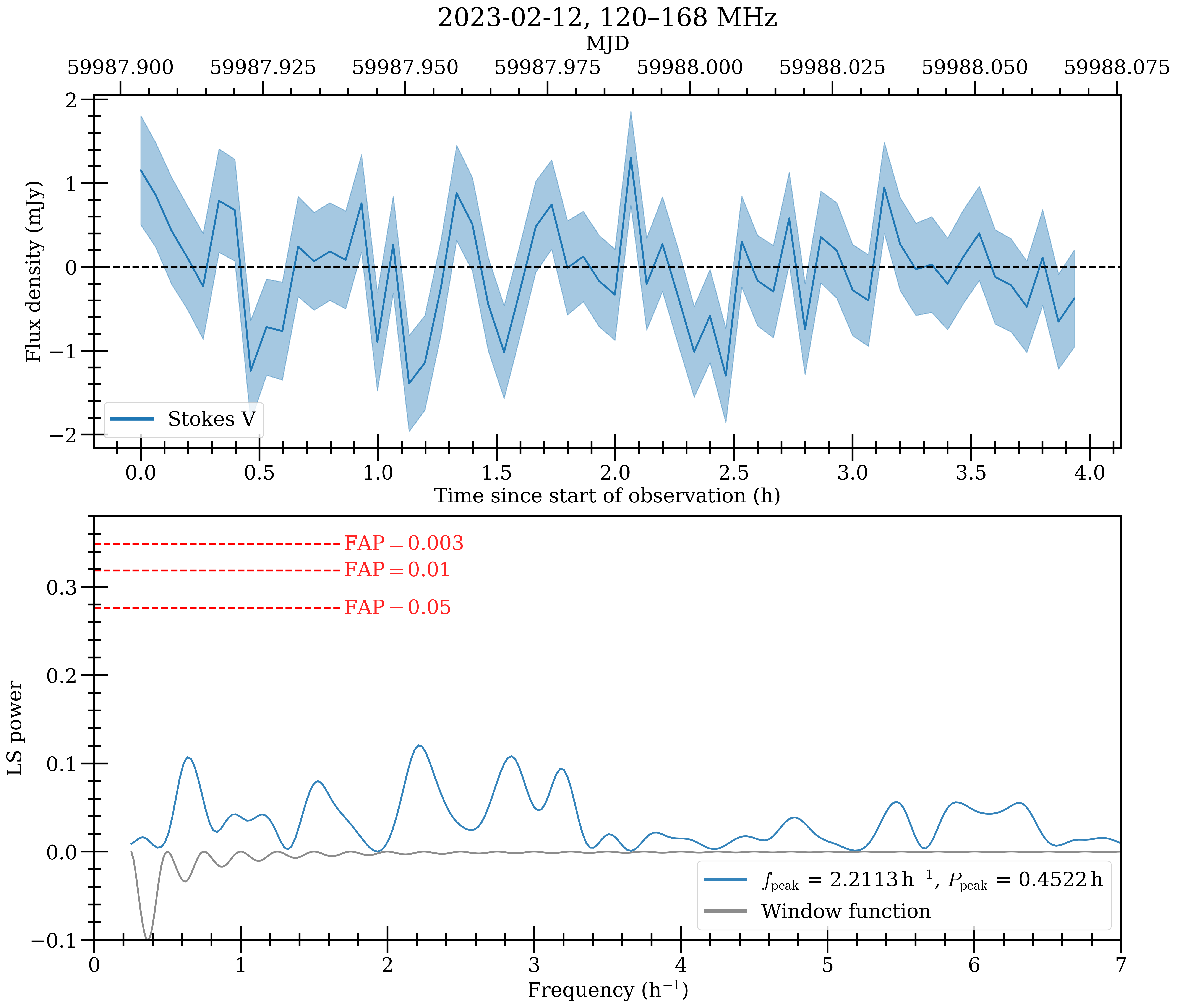}}
    \resizebox{\hsize}{!}{\includegraphics[width=0.95\textwidth]{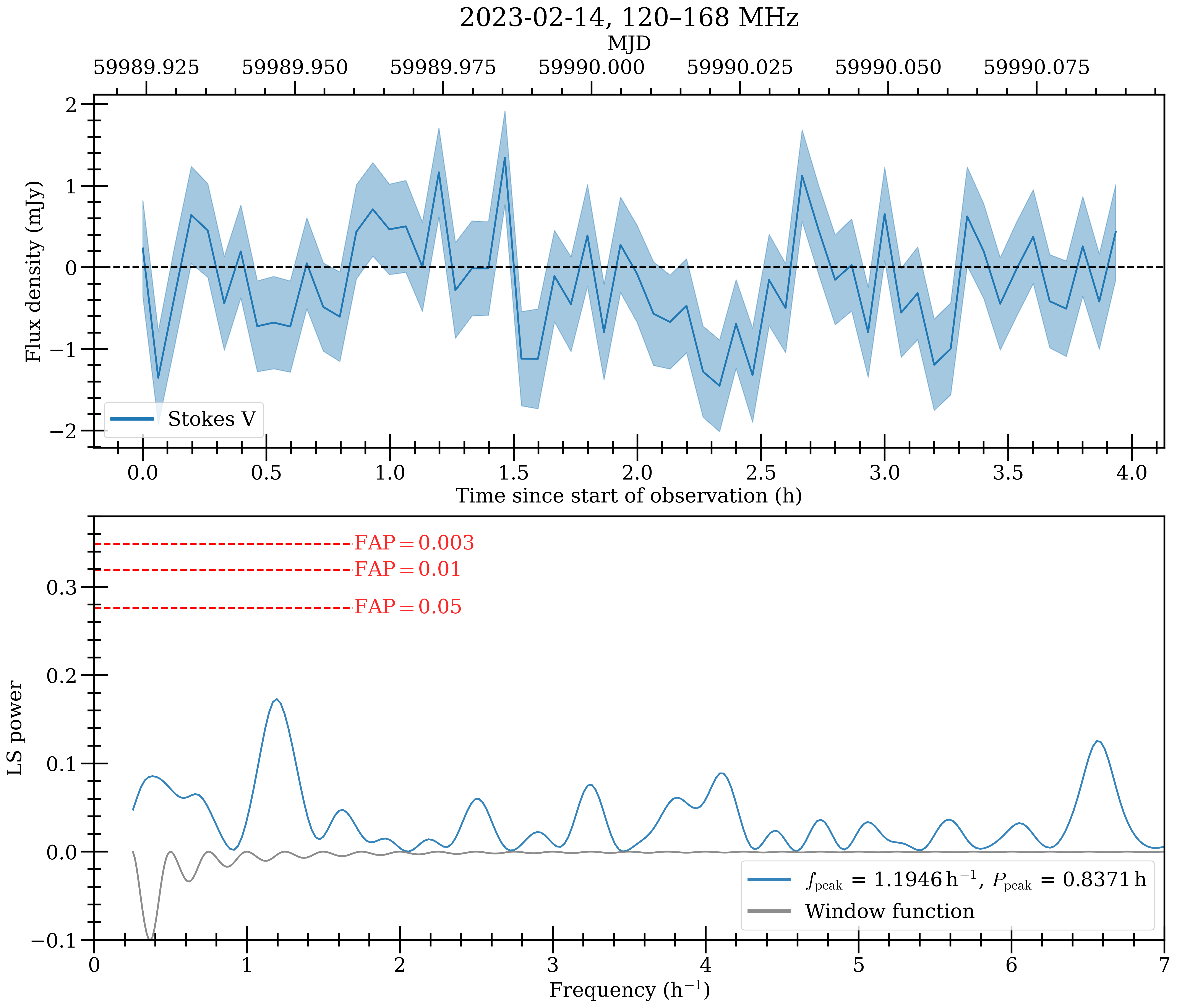}}
    \resizebox{\hsize}{!}{\includegraphics[width=0.95\textwidth]{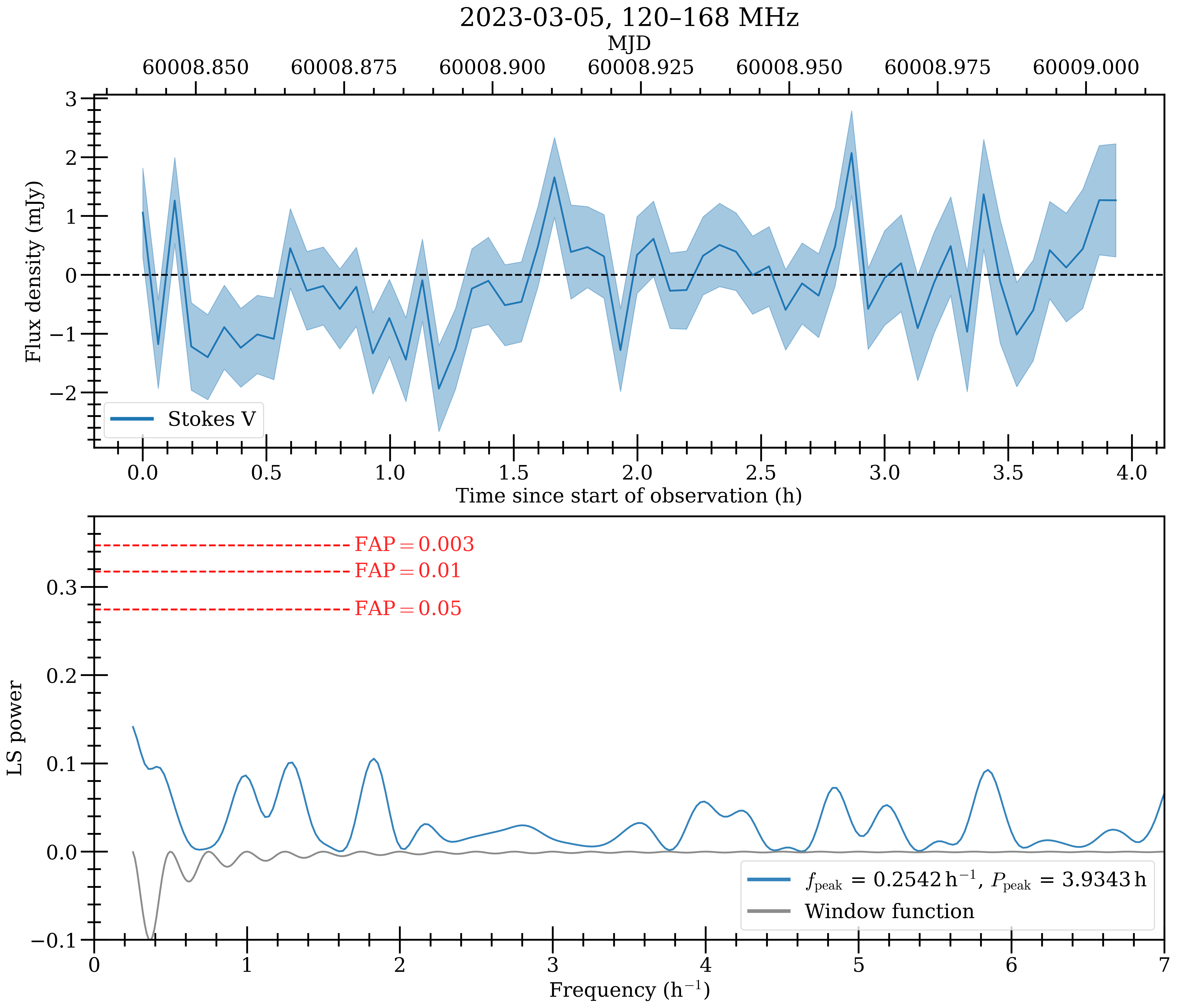}}
\end{figure}
\begin{figure}    
    \resizebox{\hsize}{!}{\includegraphics[width=0.95\textwidth]{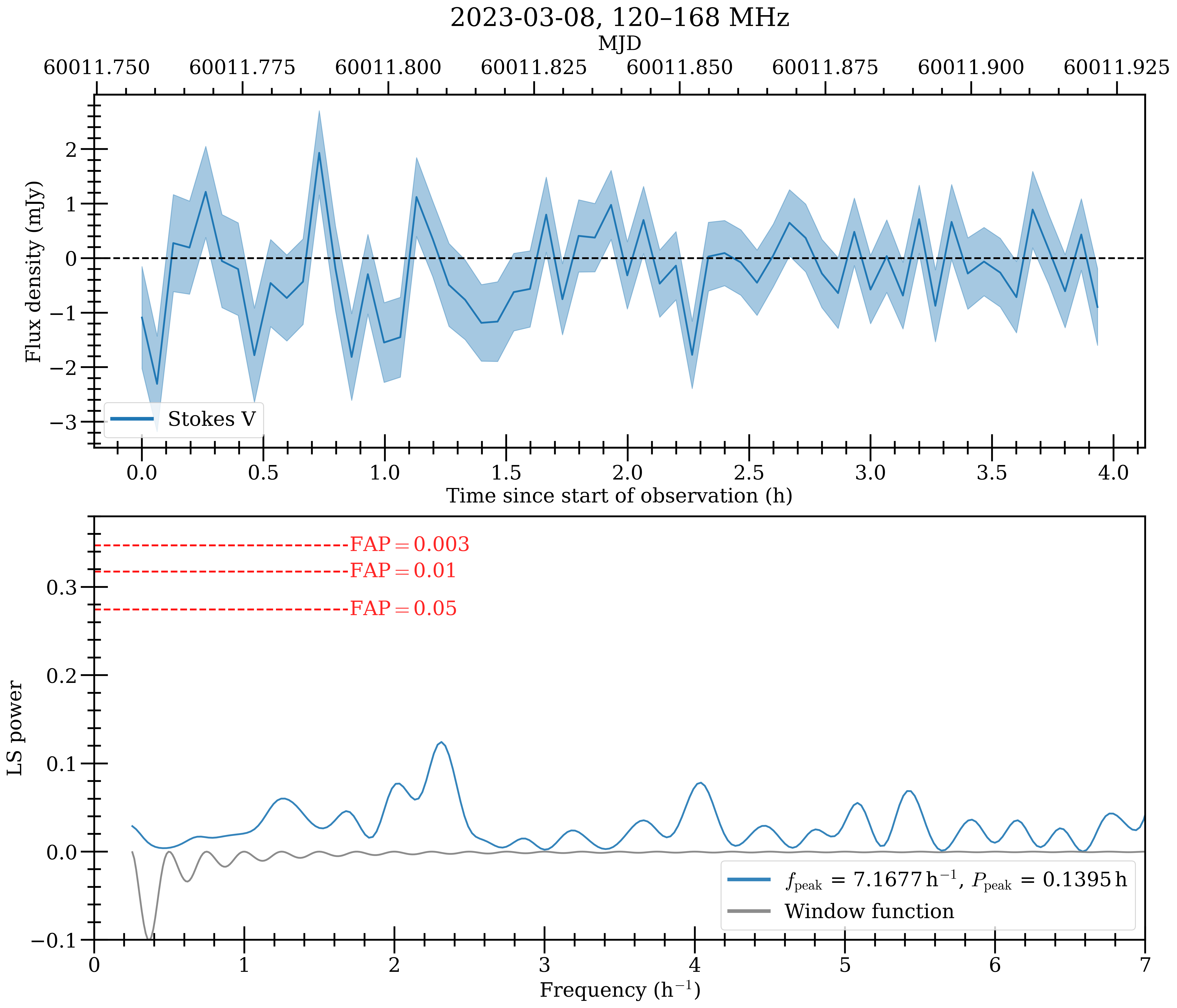}}
    \resizebox{\hsize}{!}{\includegraphics[width=0.95\textwidth]{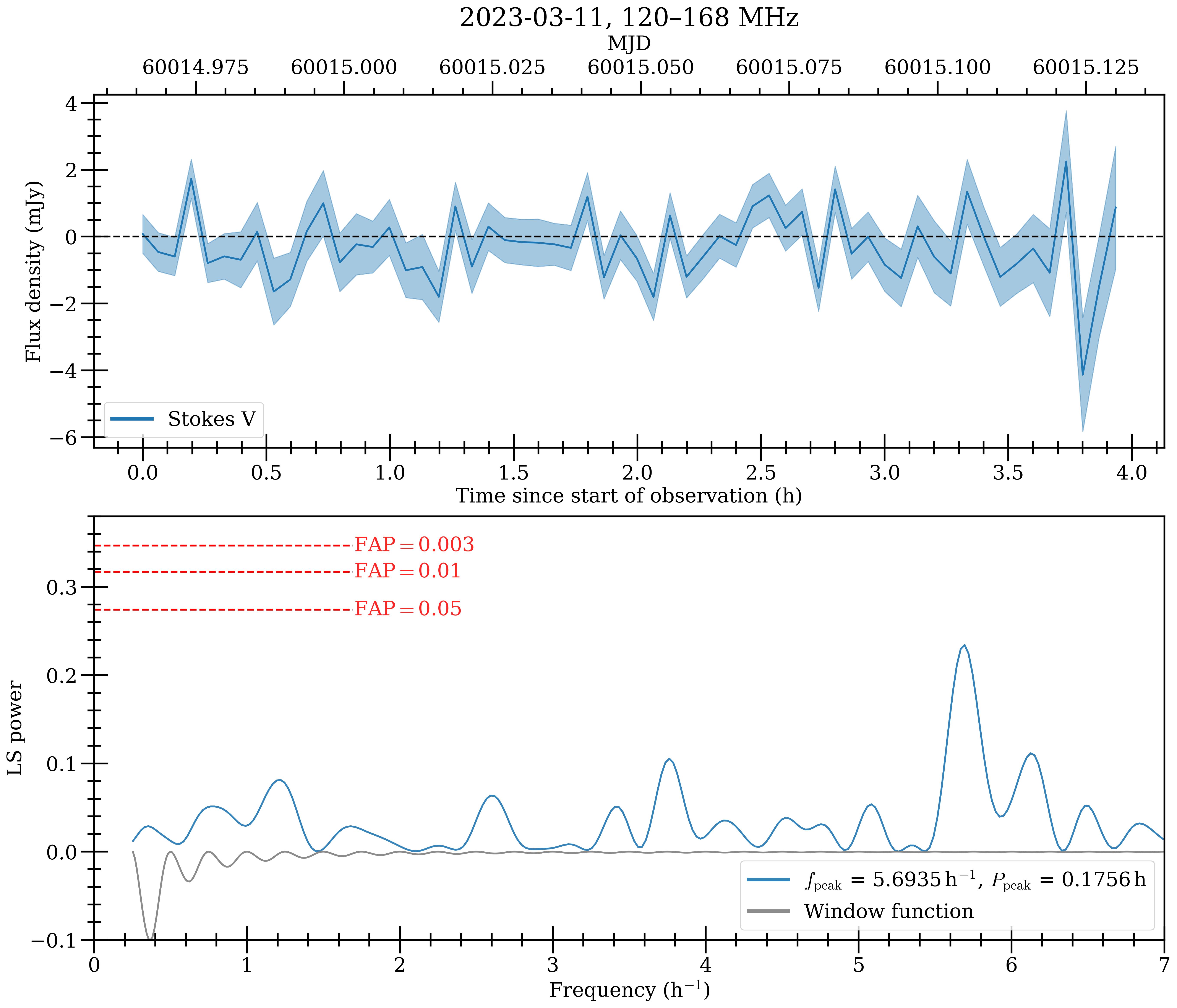}}
    \resizebox{\hsize}{!}{\includegraphics[width=0.95\textwidth]{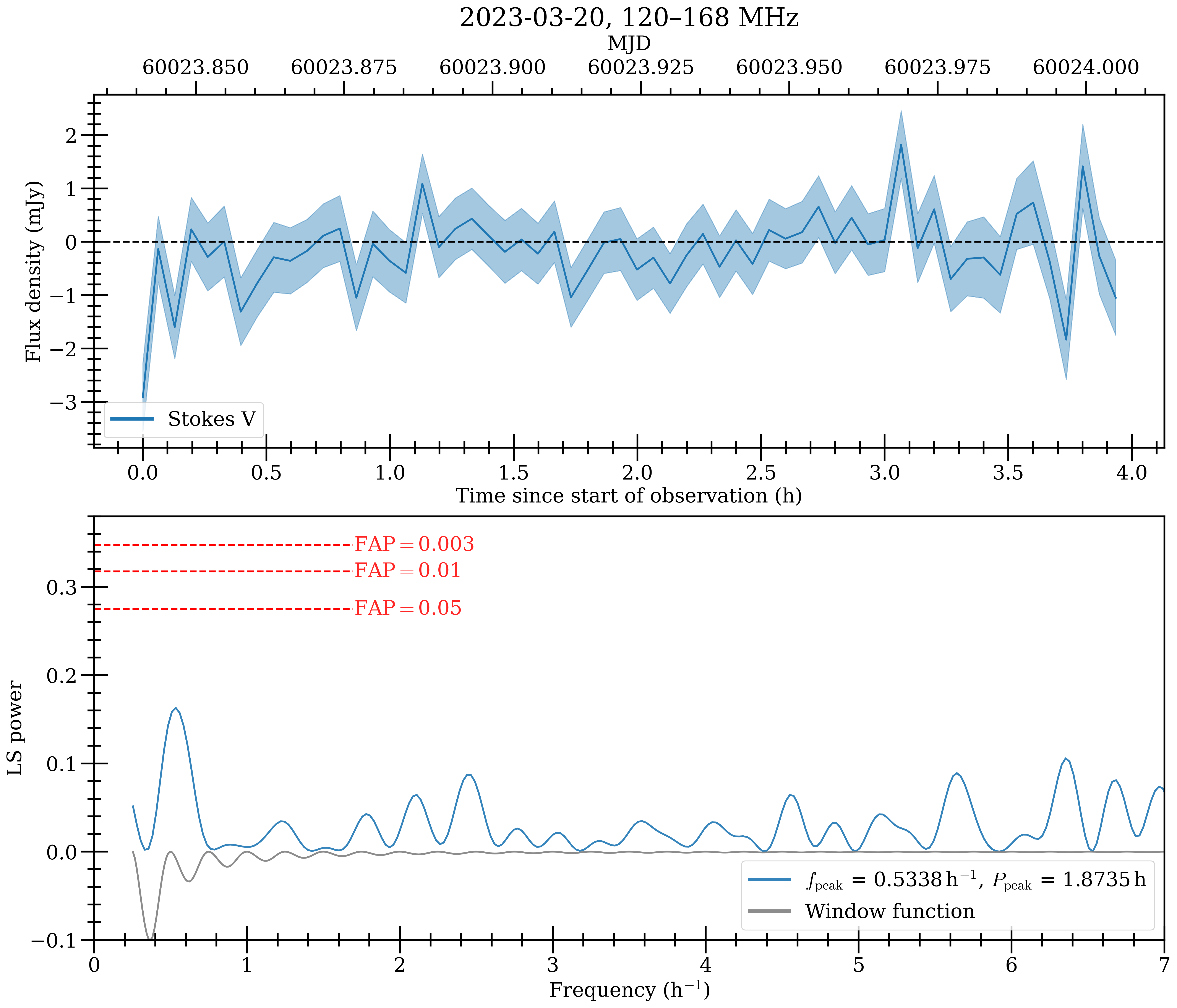}}
    \caption{Follow-up LOFAR observations of J1019+65 carried out on the remaining 6 dates.}
    \label{fig-app:lofar-6-obs}
\end{figure}

\begin{figure*}
    \centering
    \includegraphics[width=0.95\textwidth]{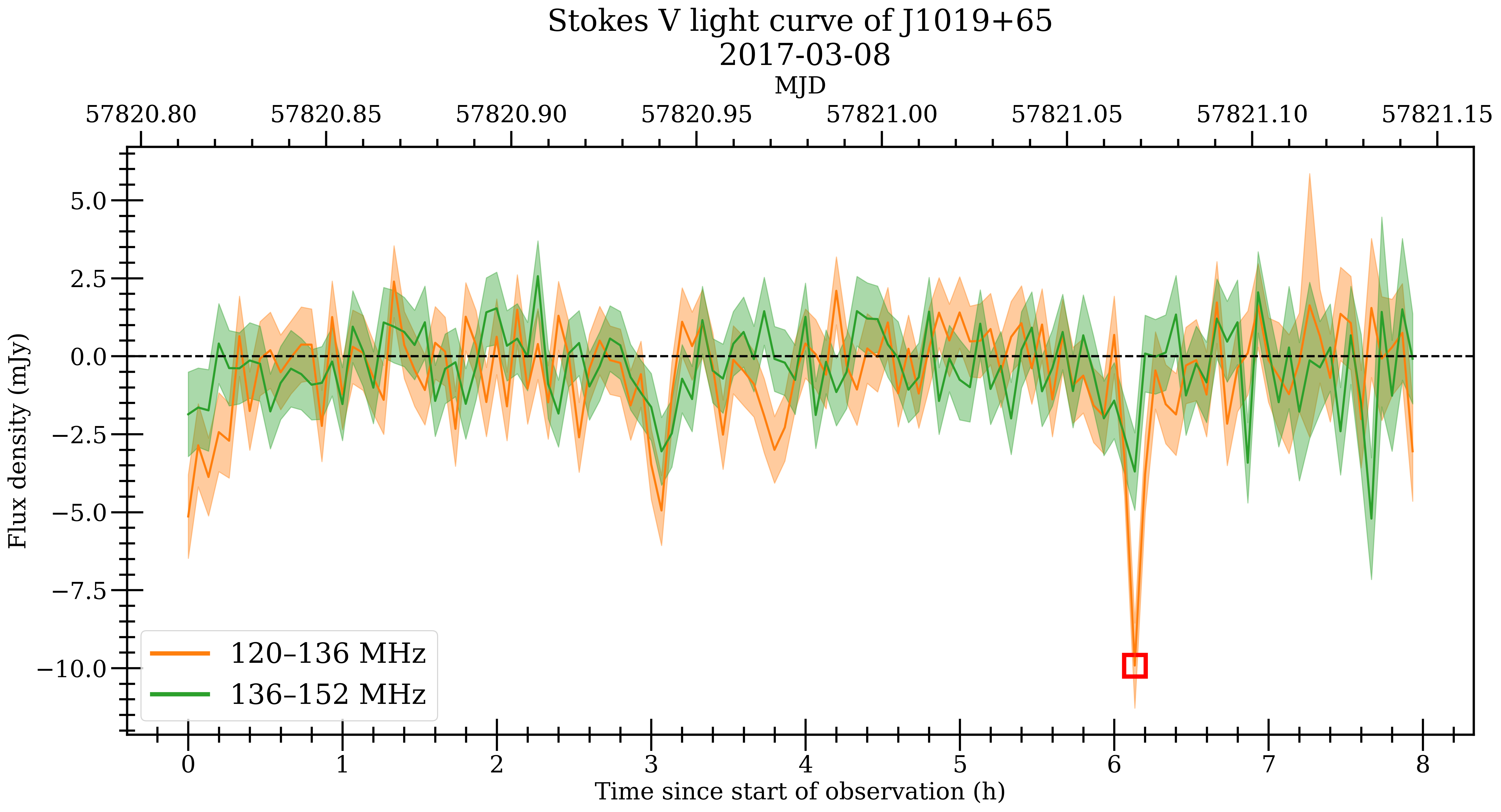}
    \includegraphics[width=0.95\textwidth]{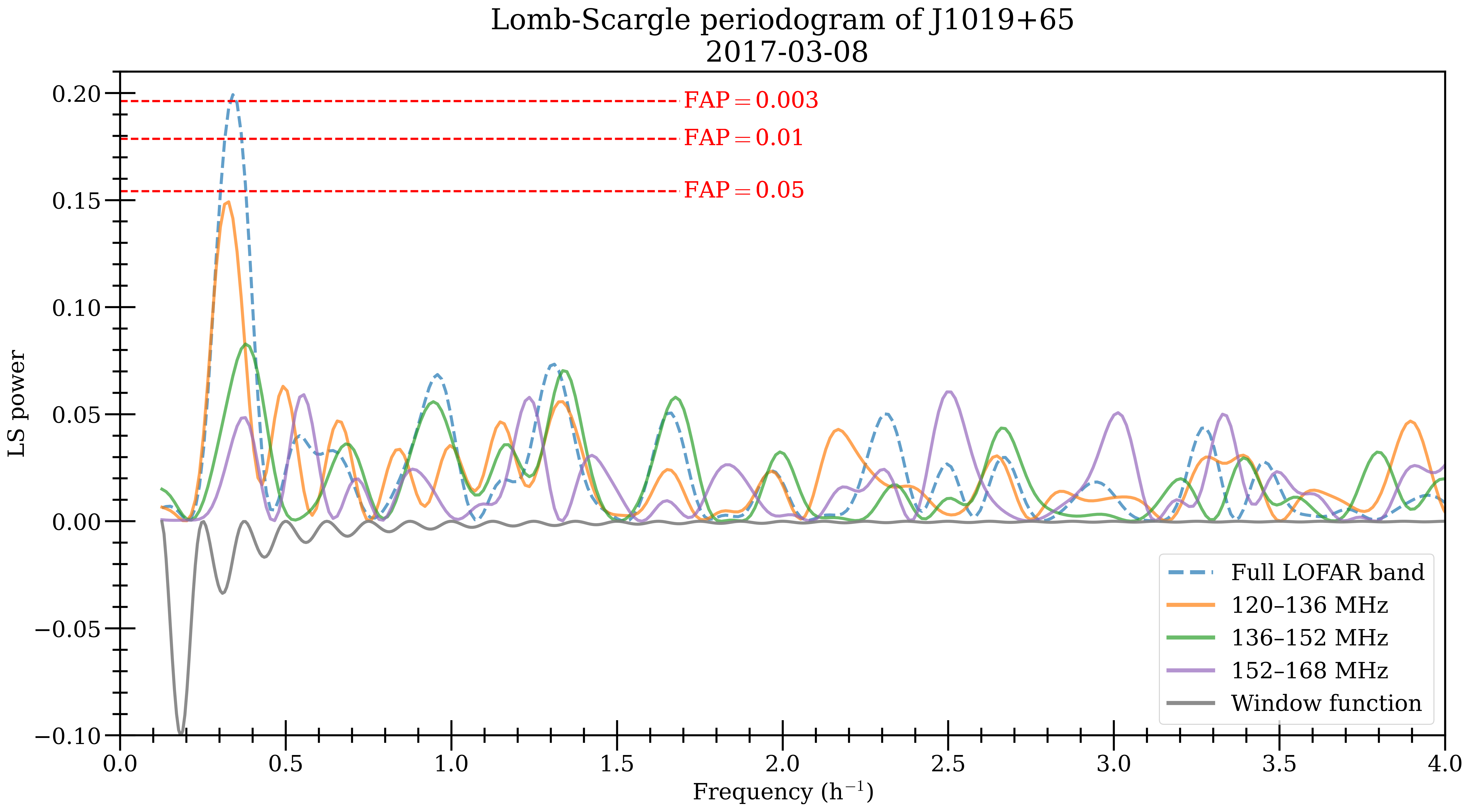}
    \caption{LOFAR observation of J1019+65 carried out on 2017-03-08 (i.e. LoTSS DR2 epoch).
    Top panel: Stokes V light curves with a cadence of 4 minutes at different frequencies. 
    The shaded region represents $\pm 1 \sigma$ uncertainty. 
    The negative $\approx 10$mJy burst at 120--136\,MHz (enclosed in red) has a signal-to-noise ratio of $> 7$. The light curve at 152--168\,MHz is omitted for the sake of clarity. 
    The top axis represents the time of observation in Modified Julian Dates (MJD). The black dashed line at 0\,mJy is drawn for clarity.
    Bottom panel: Lomb-Scargle periodogram of the radio light curves. The three red dashed lines represent the necessary LS power (i.e. peak height) to achieve a false-alarm probability (FAP) of \qty{5}{\percent}, \qty{1}{\percent}, and \qty{0.3}{\percent}. The grey curve represents the (negative) LS periodogram of the window function, which is a light curve with the same timestamps as the original curve, but the flux density of each data point is replaced with unity (i.e. a flat light curve).
    }
    \label{fig-app:lofar-2017-multi-lcls}
\end{figure*}

\begin{figure*}
    \centering
    \includegraphics[width=0.6\textwidth]{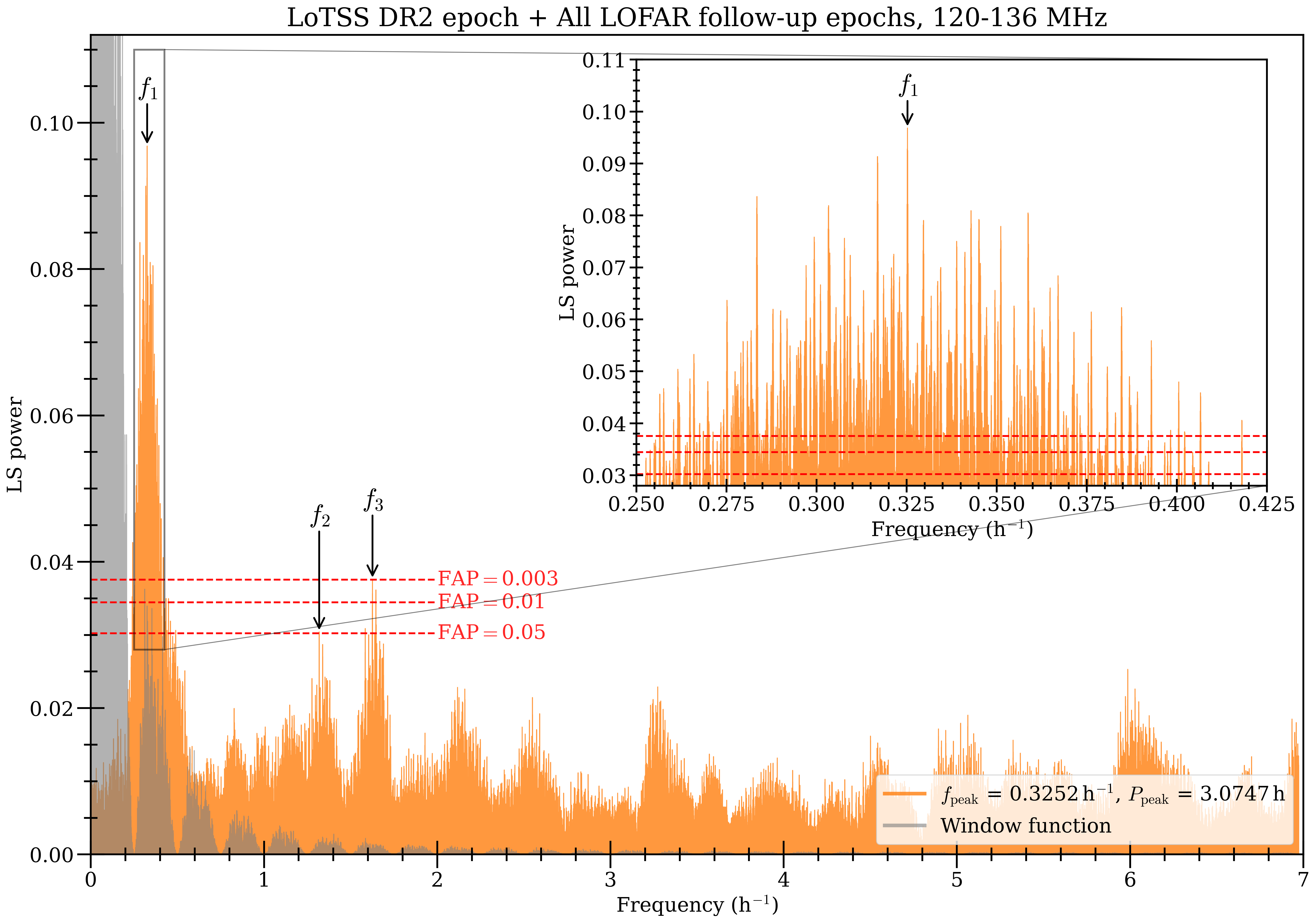}
    \includegraphics[width=0.6\textwidth]{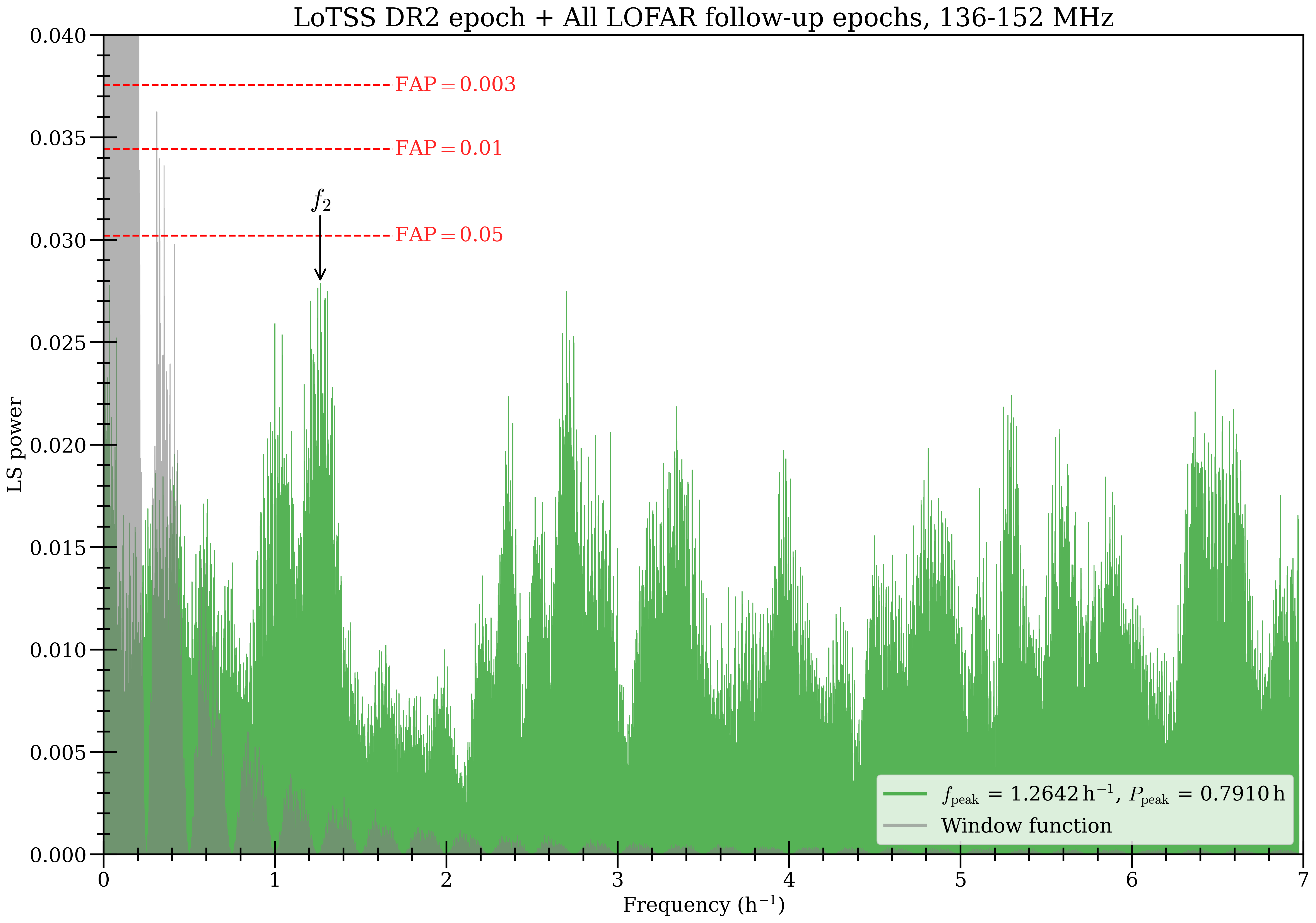}
    \includegraphics[width=0.6\textwidth]{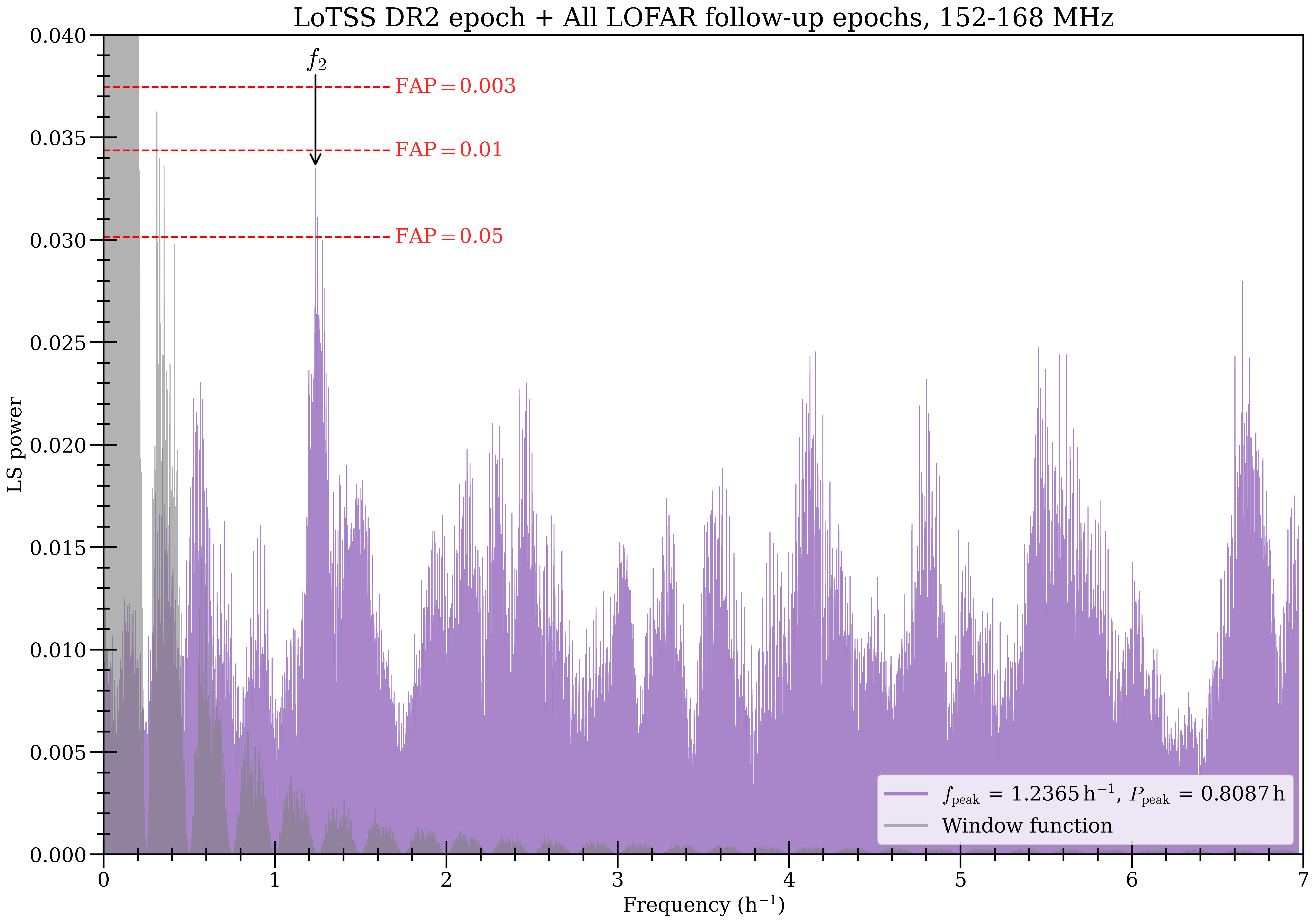}
    \caption{Lomb-Scargle periodograms of the LOFAR cross-epoch radio light curve, which includes data from both the original LoTSS DR2 epoch and our 10 LOFAR follow-up epochs.}
    \label{fig-app:cross-epoch-ls-multi}
\end{figure*}

\begin{figure*}
    \centering
    \includegraphics[width=1.0\textwidth]{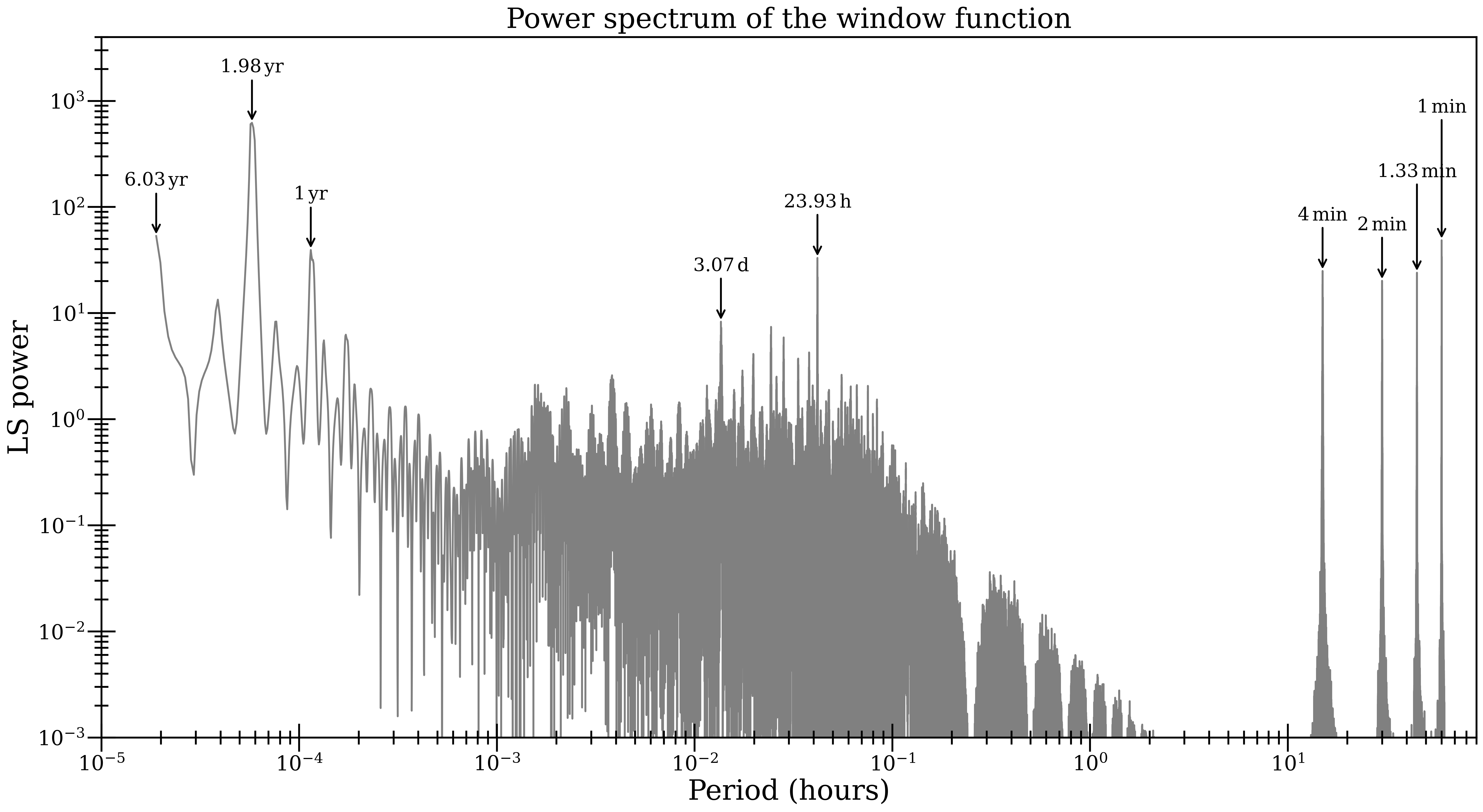}
    \caption{Power spectrum of the sampling window function for the full LOFAR dataset (cross-epoch analysis of Sec. \ref{sec:results-cross}. The light curve sampling cadence of 4min has an identifiable peak.}
    \label{fig-app:window}
\end{figure*}

\end{document}